\documentclass[sigconf]{acmart}

\AtBeginDocument{%
  }

\copyrightyear{2026}
\acmYear{2026}
\setcopyright{cc}
\setcctype{by}
\acmConference[CHI '26]{Proceedings of the 2026 CHI Conference on Human Factors in Computing Systems}{April 13--17, 2026}{Barcelona, Spain}
\acmBooktitle{Proceedings of the 2026 CHI Conference on Human Factors in Computing Systems (CHI '26), April 13--17, 2026, Barcelona, Spain}
\acmPrice{}
\acmDOI{10.1145/3772318.3791348}
\acmISBN{979-8-4007-2278-3/2026/04}

\usepackage{xcolor}
\usepackage{float}
\usepackage{multirow}
\usepackage{booktabs}
\usepackage{tabularx}
\usepackage{longtable}

\definecolor{MediaFormat}{HTML}{1B4F72}    
\definecolor{MediaElement}{HTML}{2874A6}   
\definecolor{Activity}{HTML}{1D8348}       
\definecolor{Guidance}{HTML}{239B56}       
\definecolor{Alignment}{HTML}{6C3483}      
\definecolor{Motivation}{HTML}{922B21}     

\newif\ifcomment
\commentfalse

\ifcomment
\newcommand{\hide}[1]{}
\newcommand{\note}[1]{\textcolor{blue}{<< #1 >>}}
\newcommand{\added}[1]{\textcolor{black}{#1}}
\newcommand{\newadded}[1]{\textcolor[rgb]{0.30,0.55,0.85}{#1}}

\newcommand{\cut}[1]{\textcolor[rgb]{0.5,0.5,0.5}{CUT: #1}}

\newcommand{\todo}[1]{\textcolor{red}{TODO: #1}}

\newcommand{\deleted}[1]{\textcolor[rgb]{0.8,0.8,0.8}{#1}}

\newcommand{\runze}[1]{\textcolor[rgb]{0.5,0.2,0.2}{RUNZE: #1}}
\newcommand{\zsd}[1]{\textcolor[rgb]{0.5,0.1,0.8}{ZSD: #1}}
\newcommand{\lp}[1]{\textcolor[rgb]{1.0, 0.4, 0.4}{Linping: #1}}
\newcommand{\lpr}[1]{\textcolor[rgb]{1.0, 0.4, 0.4}{#1}}
\newcommand{\Jindu}[1]{\textcolor[rgb]{1,0,0}{Jindu: #1}}

\else
\newcommand{\hide}[1]{}
\newcommand{\note}[1]{}
\newcommand{\added}[1]{#1}
\newcommand{\newadded}[1]{#1}

\newcommand{\cut}[1]{}

\newcommand{\todo}[1]{}

\newcommand{\deleted}[1]{}

\newcommand{\runze}[1]{}
\newcommand{\zsd}[1]{}
\newcommand{\lp}[1]{}
\newcommand{\lpr}[1]{}
\newcommand{\Jindu}[1]{}
\fi

\renewcommand{\quote}[1]{``\textit{#1}''}
\newcommand{\quoteby}[2]{``\textit{#2} (#1)''}

\definecolor{mypurple}{RGB}{128, 0, 255} 

\newcommand{\IBE}[0]{\textit{InteractiveBreak}}
\newcommand{\INN}[0]{\textit{Interactive Non-Narrative}}
\newcommand{\BR}[0]{\textit{Break Reminder}}
\newcommand{\VW}[0]{\textit{Video Watching}}

\begin{document}

\title[\IBE{}]{Wearable AR for Restorative Breaks: How Interactive Narrative Experiences Support Relaxation for Young \added{Adults}}

\author{Jindu Wang}
\orcid{0009-0009-4028-4662}
\affiliation{%
  \institution{Department of Computer Science and Engineering}  
  \institution{The Hong Kong University of Science and Technology}
  \city{Hong Kong}
  \country{China}
}
\email{jwangki@connect.ust.hk}

\author{Runze Cai}
\orcid{0000-0003-0974-3751}
\affiliation{%
  \institution{Synteraction Lab, School of Computing}
  \institution{National University of Singapore}
  \city{Singapore}
  \country{Singapore}
}
\email{runze.cai@u.nus.edu}

\author{Shuchang Xu}
\orcid{0000-0002-7642-9044}
\affiliation{%
\institution{Department of Computer Science and Engineering} 
  \institution{The Hong Kong University of Science and Technology}
  \city{Hong Kong}
  \country{China}
}
\email{sxuby@connect.ust.hk}

\author{Tianrui Hu}
\orcid{0009-0006-0599-3664}
\affiliation{%
\institution{School of Creative Media}
  \institution{City University of Hong Kong}
  \city{Hong Kong}
  \country{China}
}
\email{tianruhu-c@my.cityu.edu.hk}

\author{Huamin Qu}
\orcid{0000-0002-3344-9694}
\affiliation{%
\institution{Department of Computer Science and Engineering} 
  \institution{The Hong Kong University of Science and Technology}
  \city{Hong Kong}
  \country{China}
}
\email{huamin@cse.ust.hk}

\author{Shengdong Zhao}
\orcid{0000-0001-7971-3107}
\authornote{Corresponding authors.} 
\affiliation{%
\institution{School of Creative Media \& Department of Computer Science}
\institution{City University of Hong Kong}
\city{Hong Kong}
  \country{China}
}
\email{shengdong.zhao@cityu.edu.hk}

\author{Lin-Ping Yuan}
\orcid{0000-0001-6268-1583}
\authornotemark[1] 
\affiliation{%
\institution{Department of Computer Science and Engineering} 
  \institution{The Hong Kong University of Science and Technology}
  \city{Hong Kong}
  \country{China}
}
\email{yuanlp@cse.ust.hk}

\renewcommand{\shortauthors}{Wang et al.}


\begin{abstract}

Young \added{adults} often take breaks from screen-intensive work by consuming digital content on mobile phones, which undermines rest through visual fatigue and inactivity. We introduce a design framework that embeds light break activities into media content on AR smart glasses, balancing engagement and recovery, which employs three strategies: (1) seamlessly guiding users by embedding activity cues aligned with media elements; (2) transitioning to audio-centric formats to reduce visual load while sustaining immersion; and (3) structuring sessions with "rise–peak–closure" pacing for smooth transitions. In a within-subjects study (N=16) comparing passive viewing, reminder-based breaks, and non-narrative activities, \IBE{} instantiated from our framework seamlessly guided activities, sustained engagement, and enhanced break quality. These findings demonstrate wearable AR's potential to support restorative relaxation by transforming breaks into engaging, meaningful experiences.
\end{abstract}



\maketitle

\sloppy

\section{Introduction}

\begin{figure*}[ht]
    \centering
    \includegraphics[width=\textwidth]{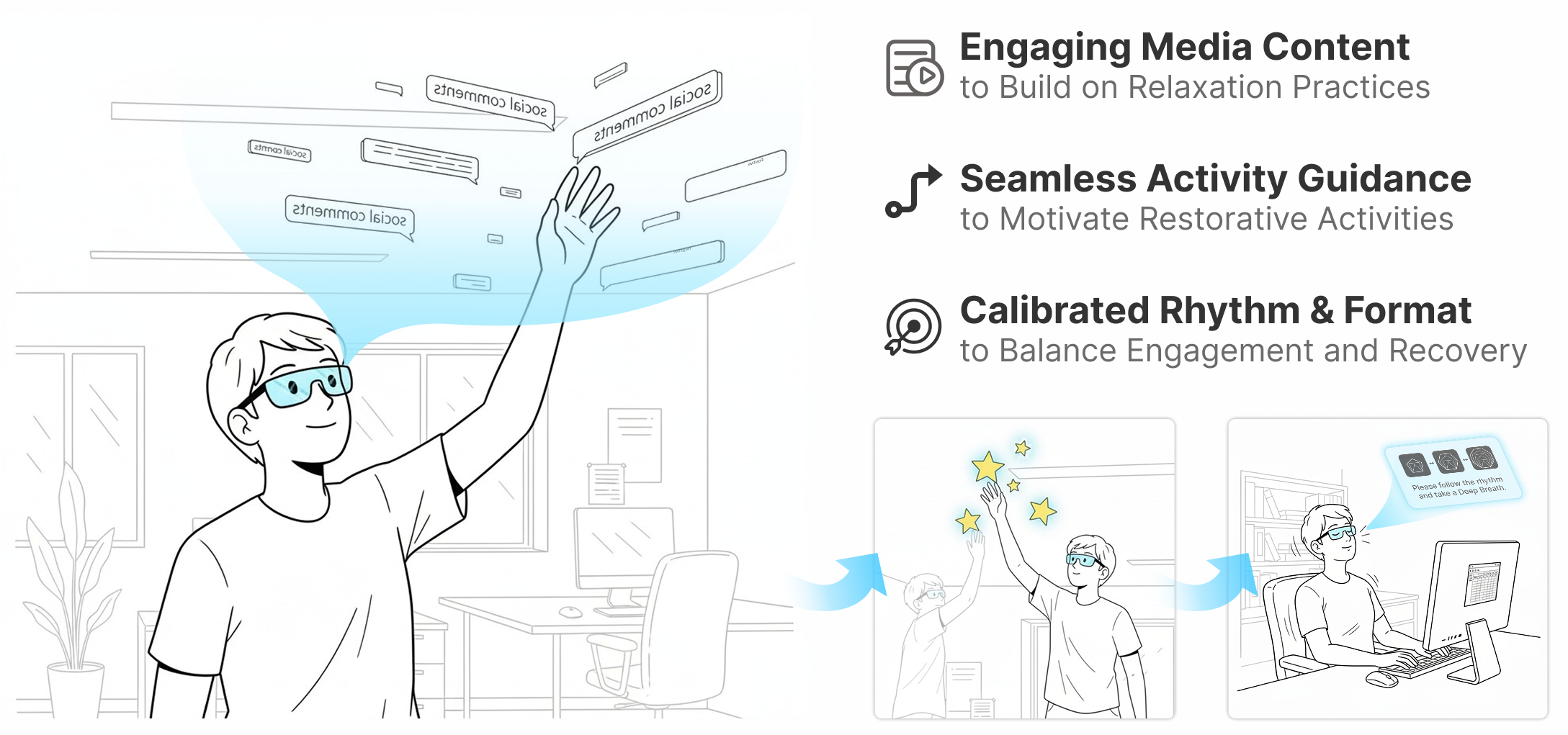}
    \caption{\newadded{Instantiated as \IBE{}, the framework enables a restorative journey where users engage with everyday audio-centric media, leveraging its intrinsic appeal to perform restorative activities seamlessly guided by cues embedded in narrative moments, orchestrated within a rise–peak–closure arc that balances immersion and recovery. Such interactive AR break experiences sustain engagement while reducing visual load to prevent over-immersion.}}
  \Description{Illustration of the InteractiveBreak framework components. The diagram highlights three design strategies: (1) ‘Engaging Media Content’ which shows a user selecting content to build on relaxation practices; (2) ‘Seamless Activity Guidance’ depicting a user performing a reaching motion guided by AR cues; and (3) ‘Calibrated Rhythm and Format’ showing a timeline of the break session transitioning from activity back to a seated work posture to balance engagement and recovery.}
  \label{fig:teaser}
\end{figure*}

\added{Knowledge workers, whose work centers on experimentation, discovery, and the generation of new insights~\cite{reinhardt2011knowledge}, often engage in prolonged and screen-intensive cognitive work. Regular breaks are therefore essential for maintaining cognitive performance and physical well-being~\cite{Epstein2016Taking5,Luo2018TimeforBreak}.} In today's digital age, young \added{adults} frequently turn to digital entertainment like videos or social media during their breaks~\cite{Tran2019PhoneChecking,xu2022typeout}. This media-driven engagement leads to inefficient, fatiguing breaks by extending sedentary visual load~\cite{Yunlong19Progress,Lee25PurposeMode}. 
However, users may still select such digital engagement for its immediacy and reward, even when they recognize the necessity of restorative breaks involving physical activities. 
A key reason for this disparity is the differing appeal of entertainment and healthy break activities.
Entertainment media captures users' attention through narrative flow, sensory rhythm, and emotional hooks, while healthy activities like stretching or mindfulness often feel effortful and under-stimulating. 
Existing approaches have attempted to encourage restorative breaks with healthy activities but have had limited success.
For example, reminder systems~\cite{Luo2018TimeforBreak,Yunlong19Progress} may not be motivating enough. Interactive and gamified approaches~\cite{Dan08SuperBreak,Scott17BreakSense,odenigbo2022journey} attempt to boost engagement, yet they struggle to compete with modern digital entertainment. Meanwhile, tools~\cite{xu2022typeout,lu2024interactout,kim2019goalkeeper} that primarily restrict digital media use focus on reducing consumption and risk overlooking the importance of providing satisfying, high-quality break experiences.

\added{Rather than removing digital media from breaks, our goal is to transform existing media habits into healthier, more restorative experiences.} We envision wearable Augmented Reality (AR) as a potential solution to promote restorative breaks by combining digital entertainment and physical activities into more engaging and rewarding experiences. By overlaying guided exercises onto real-world environments, AR can create immersive and interactive experiences that blend movement with entertainment, making physical activities feel less effortful and more stimulating. Additionally, integrating narrative flow into AR-guided breaks may provide the same sense of immersion and reward users seek from digital entertainment, while simultaneously supporting recovery and well-being. With this vision, this work explores \added{the following research question: \textit{How to design interactive AR break experiences that balance recovery support with sustained engagement, using seamless guidance mechanisms for meaningful break activities?}} 

To achieve this, we adopted a three-stage co-design process. In the first stage, we conducted semi-structured interviews with young knowledge workers to understand their expectations for meaningful yet lightweight breaks. An interesting finding was that participants preferred break experiences rooted in their habitual media behaviors rather than entirely new interactive systems. This insight suggested a promising direction: leveraging the intrinsic engagement mechanisms and content richness of entertainment media to scaffold interactive AR break experiences that seamlessly embed healthy activities into engaging media contexts. Additionally, participants emphasized the importance of seamless transitions from media to activity and of supporting restorative, socially acceptable breaks. Informed by these insights, the second stage focused on developing a video-based AR design probe~\cite{wallace2013making} that embedded activity cues into narrative flow using media elements such as character actions and spatial transitions. Based on participants’ experiences with the probe, we distilled two key design requirements: (1) achieving dual alignment between video content and physical activity to enable seamless guidance without disrupting media immersion, and (2) controlling engagement to sustain motivation without over-immersion, while maximizing the restorative quality of the break experience.
In the third stage, we conducted several design iterations through expert workshops and pilot testing, ultimately proposing a design framework to create interactive AR break experiences from videos while meeting the two requirements. 

This design framework formalizes a process for achieving dual alignment through seamless guidance design and outlines strategies across three layers: audio-centric media, activity guidance, and overall session rhythm—for controlling engagement to support high-quality recovery. 
It also outlines three phases for constructing restorative AR break experiences: (1) Contextual Scaffolding, which anchors the experience in a spatial break trajectory and audio-centric media; (2) Seamless Guidance Units, which couple narrative moments with lightweight, semantically aligned activity cues; and (3) Orchestration for Balance, which composes these units into a coherent session arc that supports both engagement and recovery.

To verify \added{the effectiveness of this framework}, we focused on three \added{evaluation aspects}: how the \added{media-embedded guidance mechanism} supports seamless transitions from media to activity, how the \added{overall break experience} balances engagement with restorative quality, and how it can adapt to different durations, contexts, and personal preferences to support long-term use. We instantiated the framework in a proof-of-concept prototype \IBE{} to operationalize these design strategies for evaluation, which transforms narrative audio content into context-aware physical activities in AR. Then a within-subjects study with 16 young knowledge workers was conducted, comparing \IBE{} against three representative break strategies: passive video watching, prompt-based reminders, and interactive non-narrative guidance. Results demonstrated \added{the promise of this media-embedded AR break approach}: \IBE{} yielded smoother transitions, higher engagement, better perceived rest quality, and stronger post-break readiness than all baselines, and showed strong potential for sustained use across diverse contexts. 

Our contributions are threefold: \textbf{(1)} We identify key user needs, expectations, and challenges in transforming media engagement into restorative AR break experiences, based on a three-stage co-design process with young knowledge workers. \textbf{(2)} We present a design framework and a proof-of-concept prototype (\IBE{}). The framework embodies strategies for \textit{seamless guidance} and \textit{calibrated engagement} to support interactive AR break experiences that are both engaging and restorative, and serves as the foundation for the \IBE{} prototype. \textbf{(3)} We contribute empirical evidence from a comparative field study, showing that \IBE{} enhances transition smoothness, engagement, and perceived restoration compared to baseline strategies, offering implications for designing interactive AR break experiences with long-term adoption potential.

\section{Related Work}
Our work extends prior research in three areas: (1) interventions for workplace well-being and break-taking, (2) transforming media into interactive and embodied experiences, and (3) augmented reality for break activity guidance.

\subsection{Interventions for Workplace Well-being and Break-Taking}

\added{To improve workplace well-being, the HCI community has focused on different aspects, including workers' stress and attention~\cite{howe2022design,sano2017designing}, digital-balance challenges~\cite{pinder2018digital}, the role of workspace and organizational factors in shaping experience~\cite{adler2022burnout,yuan2025improving}, and supporting behavioral change~\cite{lee2017self}. Among these efforts, prior work has developed several intervention mechanisms to shape workers' behaviors, motivation, and recovery~\cite{slovak2024hci,pinder2018digital,williams2023counterventions}. For example, sensing-driven approaches use physiological, behavioral, or contextual signals to surface stress or identify recovery opportunities~\cite{van2025facilitators,neupane2024momentary,jung2024deepstress,adler2022burnout}. Just-in-time systems translate these signals into micro-interventions using multimodal inputs to prompt breathing, reflection, or movement~\cite{hardeman2019systematic,suh2024toward,johnson2025adaptive,howe2022design,kaur2020optimizing}. Conversational agents and LLMs deliver lightweight micro-support through reflective dialogue, mood tracking, and mindfulness cues~\cite{yang2024exploring,yuan2025improving,mauriello2021suite}.}

\added{While these interventions identify when breaks are needed, they rarely address what happens within breaks --- how users engage and what motivates restorative activities. Many young knowledge workers spend breaks passively consuming online media~\cite{Duarte2014Web,Epstein2016Taking5}, undermining recovery through visual fatigue, inactivity, and difficult disengagement~\cite{Yunlong19Progress,Luo2018TimeforBreak,xu2022typeout}. This requires motivating meaningful action in contexts saturated with low-effort entertainment.} To this end, \textbf{reminder-based interventions}, such as time-triggered or context-sensitive prompts, have been widely explored~\cite{Luo2018TimeforBreak,Yunlong19Progress,WANG2018Persuasive}. While some systems use behavioral or physiological cues to time prompts more precisely~\cite{hardeman2019systematic}, they still struggle to integrate deeper understandings of user needs, and rarely account for the motivational gap between knowing a break is needed and wanting to do something restorative~\cite{schembre2018just}. To enhance engagement, researchers have investigated \textbf{interactive and gamified interventions} that embed play into break routines through gesture-based guidance~\cite{Dan08SuperBreak}, spatial exploration~\cite{Scott17BreakSense}, or microgames~\cite{Jiang2024PlayfulAI}. Others incorporate virtual pets~\cite{zhao2024grow,ball2021scaling} or narrative quests~\cite{murnane2020designing} to provide emotionally engaging scaffolds. Yet these systems often rely on fixed, handcrafted content and narrowly scoped mechanics. As users adapt, novelty wanes and the interactions struggle to compete with the ever-refreshing appeal of mainstream media. Another strategy, \textbf{restrictive interventions}, limits screen use via app locks~\cite{kim2019goalkeeper} or delayed unlocks~\cite{hiniker2016mytime,lu2024interactout}. While effective in curbing overuse, these approaches lack positive alternatives and may provoke resistance or disengagement by simply withholding access rather than offering value. Unlike static gamification, \textbf{media-based experiences} offer dynamic, personalized, and intrinsically engaging content formats. Our work explores \textit{how such entertainment media can be transformed into interactive break experiences} that preserve engagement while supporting physical and cognitive restoration.

\subsection{Transforming Media into Interactive and Embodied Experiences}

Research shows that traditional audio-visual media captures attention through multisensory stimulation and emotional arousal ~\cite{powell2017multimodal,sparks2013continuous}, while interactive media foster higher agency and sustained attention through bidirectional engagement ~\cite{douglas2007aesthetics,Colombati2015UserE}. Gamification strategies, including challenges, progression systems, and feedback loops—enhance engagement by appealing to intrinsic motivators like curiosity and mastery \cite{dahlstrom2012impacts,west2020challenge}. Interactive storytelling emerges as particularly compelling, embedding activities within narrative structures to scaffold engagement and support behavioral change \cite{van2009concepts,zhu2024reader,zhang2025walk}. Key mechanisms—narrative transportation, user agency, and ambiguity—foster co-creation and emotional investment \cite{cheong2006prism}. Immersive technologies like AR/VR amplify these effects through embodied experiences that enhance presence \cite{wang2024virtuwander, tian2024poeticar}, as demonstrated by applications like SnowWorld for pain distraction \cite{hoffman2020virtual} and AR-based outdoor learning systems \cite{cheng2023designing}. 
While these works demonstrate media-based interactive experiences can effectively motivate engagement and behavior, few studies have adapted these mechanisms for break requirements among young knowledge workers. Designing AR experiences that seamlessly guide users into brief, beneficial breaks remains an open challenge.

\subsection{Augmented Reality for Break Activity Guidance}

Recent augmented reality (AR) and mixed reality (MR) systems illustrate how real-time visual and auditory guidance can support a wide range of physical, cognitive, and motivational activities, offering concrete design insights for embedding restorative cues into break contexts. Visual overlays, including body outlines, trajectory traces, avatars, and target indicators, have been used to scaffold posture and movement during stretching, strength training, and aerobic tasks~\cite{diller2022visual}. Systems such as OffiStretch augment mirror views with live pose comparisons~\cite{adolf2025offistretch}, while OneBody and SkillAR project expert skeletons or movement paths in first-person perspective~\cite{Hoang2023Oneboday,diller2025skillar}, \added{demonstrating that effective guidance must align with activity structure, clearly depict goals, and enable intuitive real-time comparison and feedback.} Complementary auditory approaches reduce visual or cognitive load: AudioMove employs spatial sound to indicate limb direction~\cite{Xia2024AudioMove}, and AR running systems simulate rival footsteps or breathing rhythms to pace the user~\cite{Yuki2021AcousticAR1,Yuki2023AcousticAR2}, \added{highlighting the need for multimodal guidance that adapts to activity demands while preserving focus and environmental awareness~\cite{diller2022visual}.} Beyond physical execution, AR has also been explored for cognitive and emotional restoration; for example, Mindful Moments~\cite{Felicia23Mindful} overlays subtle cues during daily transitions to guide breathing and attention, \added{suggesting that ambient, low-effort guidance integrates well with existing routines. }Work on mixed-reality rehearsal tools~\cite{Fang2025SocialSimulation} further \added{show the importance of modulating immersion and emotional intensity while maintaining user control, especially when XR is used for short, restorative episodes rather than structured training}. Finally, AR systems have examined how to initiate and sustain beneficial practices through motivational strategies: virtual rivals, either avatars or acoustic pacers, introduce time or performance-based challenges tailored to ability~\cite{Yuki2023AcousticAR2,Hamada2022SolitaryJogging}, while narrative systems link actions to story progression or world-building, encouraging continuation by advancing characters or unlocking rewards~\cite{murnane2020designing,zhao2024grow}. \added{Building on these insights, our work examines how guidance and motivational mechanisms can be seamlessly integrated into ongoing media experiences to support lightweight, restorative break activities.}

\section{Method Overview}
\label{sec:methodoverview}

\begin{figure*}[ht]
    \centering
    \includegraphics[width=\textwidth]{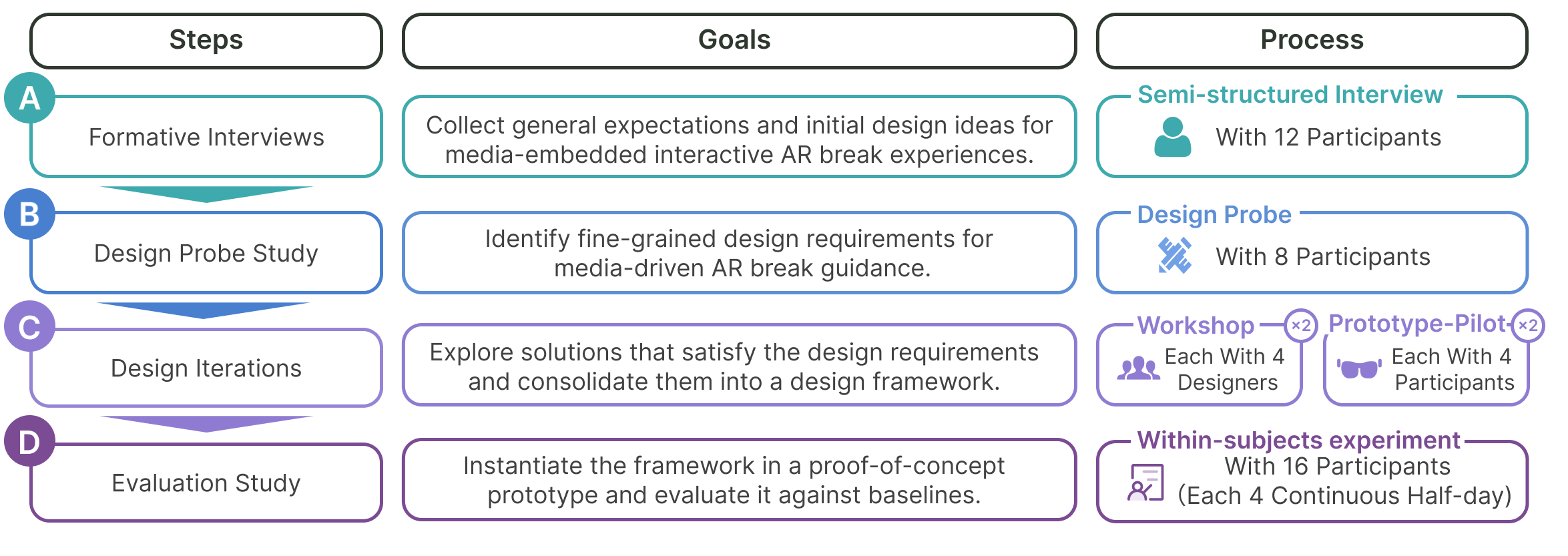}
    \Description{Flow diagram of the four-stage study design. The process flows vertically through four steps: (A) Formative Interviews with 12 participants to collect expectations; (B) Design Probe Study with 8 participants to identify requirements; (C) Design Iterations involving expert workshops and prototype-pilot cycles; and (D) Evaluation Study, a within-subjects experiment with 16 participants comparing the final prototype against baselines.}
    \caption{\added{Overview of the study design. The project used a four-part process: (A) interviews to identify expectations for interactive AR breaks, (B) a video-based AR design probe to derive fine-grained requirements, (C) iterative workshops and prototypes to develop the framework, and (D) a final comparative evaluation of the InteractiveBreak system.}}
    \label{fig:StudyWalkthrough}
\end{figure*}

\added{To investigate how AR interactive break experiences can support seamless and restorative transitions, we adopted a three-stage co-design process that progresses from expectations, to requirements, to design solutions~\cite{shi2025brickify}, followed by a final validation, as shown in Fig.~\ref{fig:StudyWalkthrough}. 
First, we conducted semi-structured interviews (N=12) to gather general expectations and initial design ideas for AR-supported interactive breaks. 
Second, we developed a video-based AR design probe and evaluated it during participants’ natural workday breaks (N=8) to test the feasibility of transforming familiar video watching into guided physical activities and to distill fine-grained design requirements. 
Third, we carried out design iterations—including two expert workshops (each N=4) and two prototype–pilot cycles (each N=4) to develop a design framework that operationalizes these requirements. 
Finally, to validate this framework, we instantiated a proof-of-concept AR prototype \IBE{}, and conducted a within-subjects evaluation (N=16) across four half-day sessions to assess its effectiveness by comparing \IBE{} with three representative baseline strategies.}

\added{All procedures were approved by the institutional review board (IRB), and informed consent was obtained from all participants. We recruited participants from a local university whose daily routines involve prolonged, visually intensive, cognitively demanding tasks spanning analytical and computational work (e.g., programming, data analysis), research and writing (e.g., literature synthesis, academic communication), and creative production (e.g., visual design, prototyping). These tasks align with widely accepted definitions of knowledge work as non-routine problem solving, information manipulation, and creative or analytical production~\cite{reinhardt2011knowledge}. Given their work characteristics and age range (20–31), our sample represents the young adult knowledge workers targeted in this study. Interviews were audio-recorded, anonymized, and transcribed. We analyzed data using Braun and Clarke's six-phase reflexive thematic analysis~\cite{braun2012thematic}. Two researchers independently coded transcripts, resolved disagreements through discussion, and collaboratively refined the codebook and themes until saturation was reached.} 

\section{Co-Design Process}
\label{sec:codesign}
\added{Building on our method overview, this section details how we progressively moved from broad user expectations to concrete design requirements and finally to design solutions. We describe three stages: (1) collecting general expectations to clarify what users want from AR-supported breaks, (2) deriving fine-grained design requirements using a video-based AR probe, and (3) iteratively developing solutions that satisfy these requirements. Each stage narrows the design space and directly informs the next, ultimately resulting in the design framework introduced in Sec.~\ref{sec:design_framework}.}

\subsection{Collecting General Expectations with Semi-structured Interviews}
\label{sec:expectations}
We first conducted semi-structured interviews with younger knowledge workers to (1) understand what expectations they may have if they can take a break with an interactive break experience, and to (2) collect their ideas on designing such interactive break experiences. 

\subsubsection{Participants and Procedure}
We recruited 12 young knowledge workers (P1–P12; 6 females, 6 males; aged 23-30) from a local university. Their daily work, including tasks such as coding, reading, and writing, required prolonged periods of sitting in front of screens, with work sessions lasting between 20 and 120 minutes. Individual interviews were conducted, each lasting approximately 40 minutes.
During the interviews, we first collected participants’ current break practices and their perspectives on effective rest. Next, we introduced potential capabilities of AR and AI, encouraging them to envision interactive, AR-supported break experiences and share their expectations for such experiences. Finally, we invited participants to sketch their design ideas for break experiences across three phases: the beginning of a break, engagement during the break, and transitioning back to work. All interviews were audio-recorded and analyzed using thematic analysis \cite{Braun01012006Thematic}.

\subsubsection{Findings: Initial Expectations and Corresponding Design Ideas.}

\added{We identified three expectations that shaped how participants envisioned meaningful interactive breaks: grounding the experience in familiar media while keeping it varied \textbf{(E1)}, enabling seamless and curiosity-driven transitions from media to activity \textbf{(E2)}, and ensuring that breaks remain lightweight, restorative, and context-appropriate \textbf{(E3)}. Table~\ref{tab:expectations} summarizes these expectations, together with brief explanations and representative participant quotes. These insights guided our interpretation of participants’ design ideas and informed the development of our design probe in the next stage.}
\begin{table*}[t]
\centering
\small
\caption{\added{Initial expectations for AR-supported interactive breaks, with descriptions and representative evidence from interviews.}}
\Description{Table summarizing three user expectations derived from interviews. The table lists: E1, the desire to build on familiar media while ensuring variation; E2, the need for seamless, curiosity-driven guidance from media to activity; and E3, the requirement for breaks to be restorative, lightweight, and context-appropriate.}
\label{tab:expectations}
\begin{tabular}{p{0.24\textwidth} p{0.36\textwidth} p{0.36\textwidth}}
\toprule
\added{\textbf{Expectation}} & \added{\textbf{Description}} & \added{\textbf{Evidence (from interview)}} \\
\midrule

\added{\textbf{E1: Build on familiar media entertainment while ensuring variation}} &
\added{Participants preferred break experiences grounded in familiar media habits (e.g., short videos) to reduce adoption effort while expecting variation to sustain interest. Videos were ideal due to their narrative continuity and rich dynamics.} &
\added{$\bullet$ \textit{"If the experience connects to the ways I usually relax, I'm more willing to adopt it... If it's entirely unfamiliar, I probably wouldn't try it."} (P2)}
\newline
\added{$\bullet$ \textit{"Even if I do the same type of break, I want it to feel different every time."} (P3)}
\newline
\added{$\bullet$ \textit{"Watching videos is already my way of relaxing—if it included healthy activities, it'd feel like a natural upgrade."} (P8)} \\

\midrule

\added{\textbf{E2: Provide seamless and curiosity-driven guidance for transitioning from entertainment to physical activities}} &
\added{Participants disliked rigid reminders and preferred transitions emerging naturally from media content. They envisioned leveraging media elements to create curiosity-driven cues that gently nudge users into movement.} &
\added{$\bullet$ \textit{"I don't like being persuaded or lectured."} (P1)}
\newline
\added{$\bullet$ \textit{"Reminders often feel stiff. I want them to be fun and closely connected to the entertainment."} (P2)}
\newline
\added{$\bullet$ \textit{"Play each piece of the video in a different spot and guide us to walk there."} (P7)} \\

\midrule

\added{\textbf{E3: Ensure restorative, lightweight, and context-appropriate breaks}} &
\added{Participants wanted breaks supporting mental and physical recovery without being socially or cognitively taxing. They emphasized a coherent break arc (initiation $\rightarrow$ activity $\rightarrow$ closure) and lightweight feedback.} &
\added{$\bullet$ \textit{"It should be effective but not tiring, and subtle enough not to draw attention in public."} (P9)}
\newline
\added{$\bullet$ \textit{"A good break includes moving my body, relaxing my mind, and then gradually calming down so I can get back to work."} (P3)}
\newline
\added{$\bullet$ \textit{"Guide me back to where I work—maybe with a character walking me there."} (P6)} \\

\bottomrule
\end{tabular}
\end{table*}

\subsection{Deriving Fine-grained Design Requirements with a Design Probe}
\label{sec:design_probe}

These findings informed our development of a video-based interactive AR break experience that transforms passive media consumption into engaging physical activities. We aimed to (1) explore whether media-driven activity guidance creates seamless transitions between content and movement, and (2) gather design requirements for AR break experiences that balance engagement with restorative quality.

\subsubsection{Design Probe}  

\begin{figure*}[t]
    \centering
    \includegraphics[width=\textwidth]{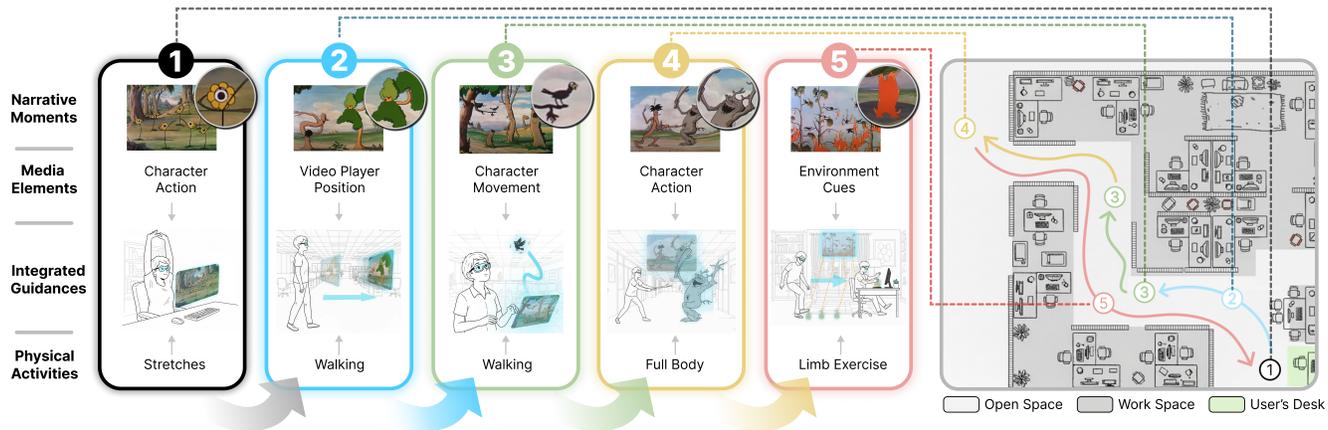}
    \caption{
    Overview of the design probe, which is a video-based interactive AR break experience set in a real workspace. It consists of five segments, each guiding users to shift their attention from a narrative moment to a physical activity, using integrated guidance drawn from media elements.}
    \Description{Five-panel storyboard of the design probe illustrating media-to-activity mappings. The diagram connects specific media elements to physical actions: (1) Character Action prompts Stretches; (2) Video Player Position prompts Walking; (3) Character Movement prompts Walking; (4) Character Action prompts Full Body exercise; and (5) Environment Cues (fire) prompts Limb Exercise. Arrows show the flow from specific narrative moments to the corresponding physical response.}
    \label{fig:probe}
\end{figure*}

We developed a \textit{video-based interactive AR break experience} that embeds physical activities into familiar media content, guided by our three expectations: building on familiar media, embedding seamless guidance, and ensuring restorative balance. 
We selected the Disney animation \textit{Flowers and Trees}\footnote{\url{https://en.wikipedia.org/wiki/Flowers_and_Trees}} as our \textbf{video media foundation (E1)}. This short film provided a clear narrative, expressive character actions, and visually distinctive scenes—rich media elements for activity integration. We then identified candidate physical activities suitable for office contexts: stretches, limb exercises, walking, and full-body movements. Designers analyzed the video to find narrative moments with salient media elements and paired them with appropriate activities, creating instances of \textbf{seamless guidance (E2)}.

As shown in Fig.~\ref{fig:probe}, each media element (e.g., character actions, video play settings, or environmental cues) was externalized into activity guidance that redirected users’ attention from the narrative to a corresponding physical activity. The probe featured five instances of seamless guidance:  
\begin{enumerate}
    \item lightweight textual prompts encouraging users to mirror a character’s motion for stretches~\cite{Hamada2022SolitaryJogging},  
    \item narrative pauses at key moments requiring users to physically relocate to resume playback,  
    \item 3D avatars moving along predefined paths to guide walking or locomotion,  
    \item monster avatars appearing during fight scenes to prompt full-body striking actions, and  
    \item fire elements that users could extinguish by stepping motions as a form of limb exercise.  
\end{enumerate}

Finally, to \textbf{structure a coherent break arc that guides users back to work (E3)}, we organized these segment--activity--guidance units into a progressive session flow. The arc began with light initiation activities, escalated into more dynamic movements, and concluded with less-intensive actions and explicit cues to return to the desk. Contextual constraints, such as available walking space and gesture visibility in shared environments, were also considered to ensure social appropriateness and workplace feasibility.

\subsubsection{Participants and Procedure}  
We assessed the probe with 8 young knowledge workers (5 female, 3 male; aged 22--28; distinct from the initial interview group) during their natural workday breaks. Each participant engaged in a 10-minute break session after a 40-minute work period, followed by a 30-minute semi-structured interview. The interviews explored experiences, preferences, and contextual considerations regarding the media-driven activity guidance and overall AR break experience, aiming to identify actionable design requirements, with  \added{codebook provided in Appendix~\ref{sec:appendix_codesign}.}

\subsubsection{User Feedback on the Probe}
Overall, participants noted that our media-driven activity guidance and the entire experience were effective. Their feedback mainly fell into two categories. 

\textbf{First, media-driven activity guidance can encourage physical activities because of its natural integration of the activities and narrative flow.} 
Media-driven activity guidance made participants feel natural rather than prescribed, since the guidance emerged organically from the narrative flow. For example, P6 said, \quote{The transitions were smooth; it felt like the story moved into the environment and pulled me along}. Participants particularly enjoyed embodied interactions, such as fighting or extinguishing fire, which were described as \textit{"fun," "interactive,"} and \textit{"intuitively connected to what was happening in the video."} 

\textbf{Second, the whole experience is effective for structured rest because it combines mental lightness with physical activation.}
Many participants valued that the experience subtly encouraged movement without feeling demanding, with one noting, \quoteby{P3}{Normally I just sit and watch, but this made me get up and move without realizing, which was satisfying}. Several participants reported feeling \textit{"refreshed," "more awake,"} and \textit{"ready to resume work,"} suggesting that the experience could successfully balance engagement with recovery. The sense of variety and novelty was also highly valued, as P6 stated, \quote{It felt like a short adventure --- each time different and always worth it}.

\subsubsection{Design Challenges and Design Requirements} Despite the promising outcomes of the probe, it also surfaced two key design tensions that require careful consideration, leading to two corresponding design requirements. 
  
\textbf{C1: Aligning guidance with media and context to avoid interruption.}
The first challenge lies in ensuring guidance is both semantically tied to the unfolding media content and actionable within the user’s real-world context, preserving the continuity of the media experience.
Participants’ accounts revealed what makes guidance feel continuous and uninterrupted.
Specifically, on the \textbf{media side}, they emphasized that guidance must be closely tied to the narrative. As P6 explained, \textit{“I was willing to follow because the guidance felt naturally connected to what was happening in the video”}, while P7 cautioned, \textit{“If there is no pull from the storyline, I won’t enter the next activity node.”} Similarly, P4 noted, \textit{“The tighter the link with the plot, the better the activity feels.”} On the \textbf{activity side}, participants stressed that actions had to remain light, clear, and immediately feasible in context; otherwise, immersion would break. As P4 remarked, \textit{“If the activity doesn’t fit the physical or social environment, I immediately feel pulled out of the experience.”} 

\textbf{DR1: Stay “inside” the media through dual alignment with content and activity.}  
To address the challenge \textbf {(C1)}, we articulate the principle of \textbf{dual alignment}: cues should be semantically embedded in the narrative while also being light, providing clear feedback, and contextually feasible. This ensures that even when users are prompted to act, they still perceive themselves as \emph{inside the media experience} rather than switching tasks. By clarifying dual alignment as the defining characteristic of seamless guidance, designers can move beyond fragmented examples and establish a reproducible method for constructing seamless guidance across varied media and activity contexts.  

\textbf{C2: Fostering visual rest and clear break boundaries.}  
Participants described the probe as highly immersive and motivating, but this strength also created tensions. Some felt reluctant to end the session, saying they \textit{“wanted to keep playing”} (P5) or continued replaying the experience after sitting back down (P6). Others noted that continuous visual attention provided little opportunity for genuine eye rest, especially for those already experiencing screen fatigue (P4, P5, P7). These accounts revealed that while engagement was valuable, it sometimes tipped into over-immersion, blurring the boundary between rest and work and undermining recovery.  

\textbf{DR2: Controlling engagement to be motivating yet not over-immersive to maximize break quality.}  
To resolve the tension \textbf{(C2)}, we propose that engagement should be deliberately calibrated rather than maximized. Designers should harness media’s immersive qualities to motivate activity, but regulate intensity, duration, and visual demand so that the experience sustains restorative actions without overloading attention. The ideal design provides just enough engagement to initiate and carry users through activities, while preserving opportunities for visual rest and ensuring a clear sense of closure that supports readiness to return to work, thereby maximizing the restorative quality of breaks.

\subsection{Finding Solutions to Meet Design Requirements with Design Iterations}
\label{sec:design_iterations}

In order to address the two core design requirements (i.e., \textbf{DR1} and \textbf{DR2}), we conducted a series of design iterations. 
Specifically, we conducted two workshops (each with 4 participants from HCI, AR/VR design, film and media production, industrial design, and communication design) and two rounds of prototype–pilot cycles (each with 4 distinct participants). In each workshop, designers first created instances of guidance by mapping media moments to activity cues that satisfied the requirements of seamlessness (exploring DR1), and then refined and composed these into complete interactive break experiences that calibrated engagement to maximize restorative quality (exploring DR2). Prototypes were then implemented and piloted, progressively shifting formats and guidance strategies based on findings. This cycle of design exploration, prototyping, pilot validation, and redesign generated the insights that informed our key design decisions. 
After these iterations, we consolidated the design knowledge into a design framework, as elaborated in Sec.~\ref{sec:design_framework}.

\section{Design Framework}
\label{sec:design_framework}

We propose a design framework for transforming engaging video or audio media into restorative AR break experiences. \added{At its core, an AR interactive break consists of lightweight, media-integrated guidance, each prompting a small restorative activity, positioned within a spatial and temporal scaffold that defines where actions can occur, and arranged into a coherent session arc that balances engagement with recovery.} \added{In the following sections, we describe how to construct each sub-task through our \textbf{seamless guidance design considerations (DR1)}, which ensure that guidance emerges naturally from the ongoing media experience, and how to coordinate activity guidance instances using our \textbf{engagement–restoration strategies (DR2)}, which shape the overall session arc to sustain motivation while supporting cognitive and physical recovery. Together with the contextual scaffold, these components form a complete design framework for creating restorative, media-embedded AR break experiences.}

\subsection{Seamless Guidance Design Considerations (DR1)}
\label{sec:guidance_design}

\begin{figure*}[htbp]
    \centering
    \includegraphics[width=\textwidth]{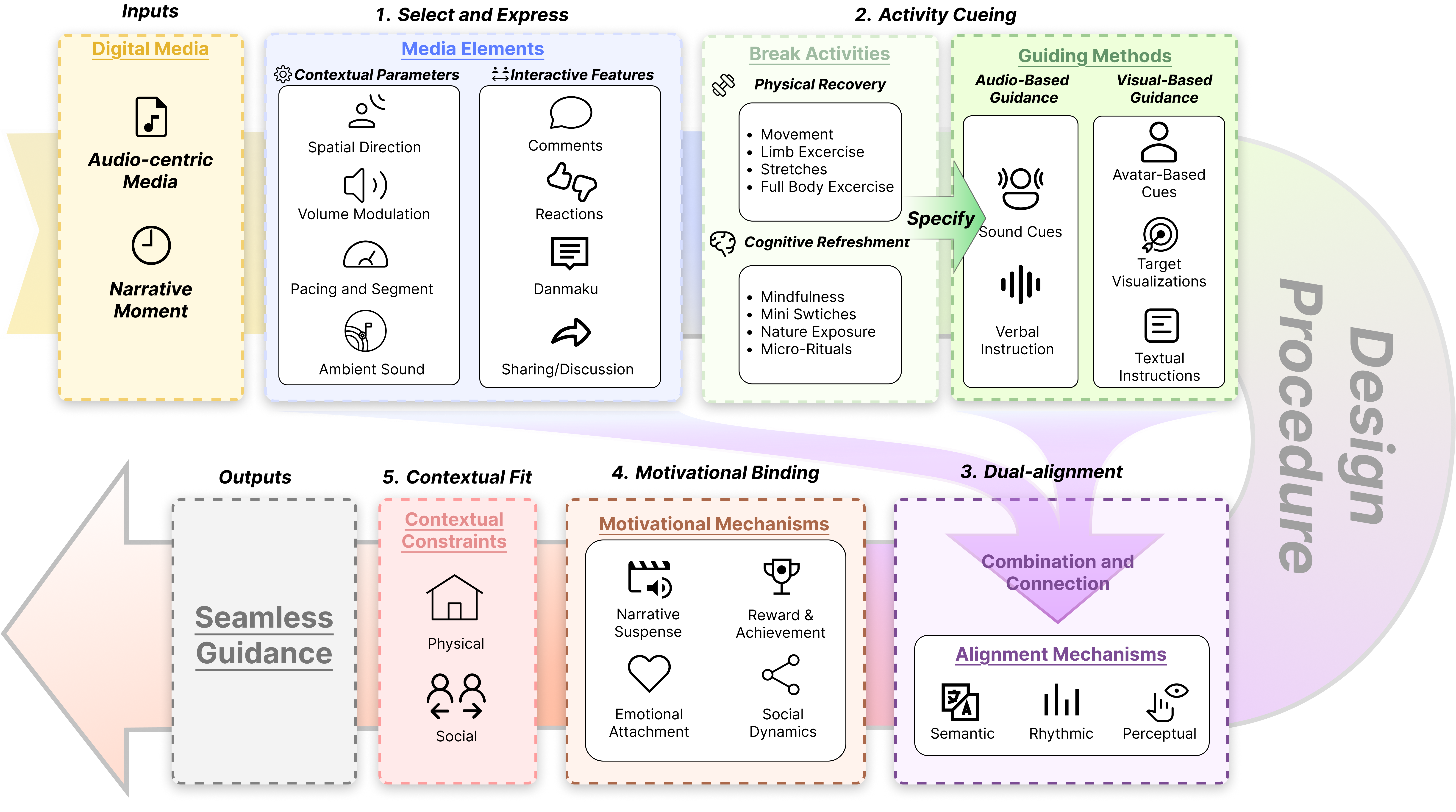}
    \caption{\added{Design process of Seamless Guidance Unit: Seven components: Digital Media, Break Activities, Media Elements, Activity Cues, Alignment Dimensions, Motivational Mechanisms, and Contextual Constraints, informing the Seamless Guidance Unit (SGU) and its design procedure.}} 
    \Description{Schematic of the Seamless Guidance Unit (SGU) design process. The diagram flows from left to right, starting with Inputs (Digital Media, Audio-centric Media, Narrative Moment). It passes through five processing steps: (1) Select and Express Media Elements; (2) Activity Cueing (Audio/Visual); (3) Dual-alignment (Semantic, Rhythmic, Perceptual); (4) Motivational Binding; and (5) Contextual Fit (Physical/Social constraints). The final output is the Seamless Guidance unit.}
    \label{fig:framework}
\end{figure*}

\added{To address DR1 of aligning media content and activity, we distilled a set of interconnected design considerations from iterative prototyping and participant feedback. These considerations characterize how narrative moments, media expressions, cue modalities, motivational levers, and contextual constraints jointly support seamless transitions into restorative action. Table~\ref{tab:sgu_framework} summarizes the core considerations for constructing a Seamless Guidance Unit (SGU), outlining the key components involved and the five-step procedure that translates narrative moments into lightweight, media-integrated guidance. The table specifies what designers can consider at each stage—from externalizing media moments and cueing activities to ensuring alignment, motivation, and contextual fit—and illustrates how these elements work together to preserve immersion. Fig.~\ref{fig:framework} provides a complementary visual overview of the structure, depicting how the seven components map onto the SGU design steps and jointly support seamless transitions into restorative action.}

\begin{table*}[htbp]
\centering

\renewcommand{\arraystretch}{1.0} 

\setlength{\tabcolsep}{3pt}

\footnotesize 

\caption{\added{Seamless Guidance Unit (SGU) design framework and procedure.}}
\Description{Table detailing the Seamless Guidance Unit design framework. It organizes the design process into specific stages (Inputs, Media Externalization, Activity Cueing, Dual Alignment, Motivational Binding, Contextual Fit, and Outputs), listing the criteria and component examples (such as specific audio cues or visual targets) for each stage.}
\label{tab:sgu_framework}

\begin{tabular}{@{} p{0.12\textwidth} p{0.40\textwidth} p{0.44\textwidth} @{}}
\toprule
\textbf{\added{Design Stage}} & \textbf{\added{Action \& Criteria}} & \textbf{\added{Component Focus \& Examples}} \\
\midrule
\textbf{\added{Inputs}} & 
\added{Select a \textit{Narrative Moment} (brief temporal slack/semantic leverage) and a target \textit{Break Activity} (lightweight/restorative).} & 
\added{\textbf{1. Digital Media:} Audio-centric (podcasts); focus on narrative moments like topic shifts~\cite{Xu_2025}. \newline \textbf{2. Break Activities:} Physical (stretching) \cite{10.1145/3715070.3749268}, cognitive (mindfulness).} \\ 
\midrule
\textbf{\added{Step 1:}\newline \added{Media Externalization}} & 
\added{\textbf{Select and express the media element.} \newline \textit{Criterion:} High recognizability, low visual load, minimal narrative disruption.} & 
\added{\textbf{3. Media Elements:} \newline $\bullet$ \textit{Contextual Params:} Spatial audio drift, volume modulation, pacing. \newline $\bullet$ \textit{Embedded Features:} Reactions, comments, danmaku~\cite{Xu_2025,douglas2007aesthetics,Colombati2015UserE}.} \\ 
\midrule
\textbf{\added{Step 2:}\newline \added{Activity Cueing}} & 
\added{\textbf{Specify activity cueing.} \newline \textit{Criterion:} Make action immediately clear (what/how) with low salience.} & 
\added{\textbf{4. Activity Cues:} \newline $\bullet$ \textit{Audio-based:} Spatial sounds, rhythmic pacing, verbal instructions~\cite{Xia2024AudioMove,Yuki2021AcousticAR1}. \newline $\bullet$ \textit{Visual-based:} Avatar demos, target visualizations~\cite{Dan08SuperBreak,Jiang2024PlayfulAI}.} \\ 
\midrule
\textbf{\added{Step 3:}\newline \added{Dual Alignment}} & 
\added{\textbf{Apply alignment dimensions.} \newline \textit{Criterion:} Ensure the transition feels like a media continuation, not a disruption.} & 
\added{\textbf{5. Alignment Dimensions:} \newline $\bullet$ \textit{Semantic:} Cue meaning resonates with narrative~\cite{cheng2023designing}. \newline $\bullet$ \textit{Rhythmic:} Syncs with pacing/beats~\cite{10.1145/3746059.3747613,10.1145/3746059.3747791}. \newline $\bullet$ \textit{Perceptual:} Matches active sensory channel~\cite{odenigbo2022journey}.} \\ 
\midrule
\textbf{\added{Step 4:}\newline \added{Motivational Binding}} & 
\added{\textbf{Add motivational binding.} \newline \textit{Criterion:} Sustain participation without diverting attention from recovery.} & 
\added{\textbf{6. Motivational Mechanisms:} \newline $\bullet$ Narrative suspense, emotional attachment~\cite{murnane2020designing, zhao2024grow,10.1145/3654777.3676336}. \newline $\bullet$ Rewards, social dynamics (reactions/comparisons)~\cite{Scott17BreakSense,Yuki2023AcousticAR2}.} \\ 
\midrule
\textbf{\added{Step 5:}\newline \added{Contextual Fit}} & 
\added{\textbf{Conduct safety/fit pass.} \newline \textit{Criterion:} ``Actionable in context'' regarding space, norms, and time.} & 
\added{\textbf{7. Contextual Constraints:} \newline $\bullet$ \textit{Physical:} Safety, walkable paths~\cite{10.1145/3411814,10.1145/3463524}. \newline $\bullet$ \textit{Social:} Visibility, noise norms. \newline $\bullet$ \textit{Time:} Feasibility within budget~\cite{Luo2018TimeforBreak}.} \\ 
\midrule
\textbf{\added{Outputs}} & 
\added{\textbf{Final SGU Specification:} Media element, cueing details, alignment rationale, motivational binding, and contextual adjustments (with duration/intensity).} & 
\added{--} \\ 
\bottomrule
\end{tabular}
\end{table*}

\subsection{Design Strategies for Balanced Break Quality} 
\label{sec:balance_strategies}
\begin{figure}[htbp]
    \centering
    \includegraphics[width=0.5\textwidth]{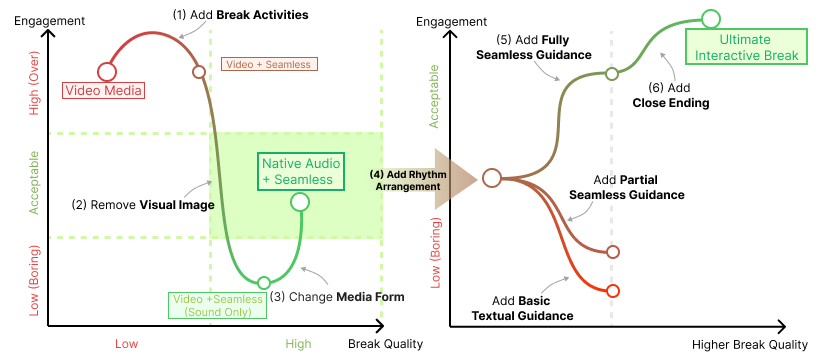}
    \caption{Iteration process for balancing engagement and recovery:  
(1) add seamless activity guidance into video media (design probe);  
(2) remove video visuals to reduce visual load;  
(3) shift to native audio media as the format for seamless guidance;  
(4) structure the break as an arc—light start, engaging middle, calming closure;  
(5) pilot study confirms all activities should remain seamlessly media-embedded;  
(6) add a dedicated close ending to reinforce completion.}
    \Description{Line graph plotting Engagement (Y-axis) against Break Quality (X-axis) across six design iterations. The line traces a path starting at high engagement but low break quality (Video Media). It dips in engagement when visuals are removed, then rises in both metrics as Audio-centric media is introduced. The highest point in both Engagement and Break Quality is reached at step 6 (Interactive Break), which incorporates 'fully seamless guidance' and a 'close ending' structure.}

    \label{fig:iteration}
\end{figure}

To meet \textbf{DR2}, as shown in Fig.~\ref{fig:iteration}, we iterated on three aspects of the experience: \emph{media choice}, \emph{activity guidance}, and \emph{overall structure}.
This process yielded two strategies for calibrating engagement and maximizing break quality.

\textbf{Controlling media immersion while enhancing cognitive recovery through audio-centric content.}
Early probes using video as the media foundation (Fig.~\ref{fig:iteration}-(1)) proved highly engaging but difficult to disengage from, with participants reporting lingering attachment and excessive visual demand that limited eye rest. Removing visual components (Fig.~\ref{fig:iteration}-(2)) reduced fatigue but created new problems: without visual scaffolding, participants struggled to follow stories and experienced increased cognitive load. This led us to shift from deconstructed video to inherently audio-native media such as talk shows, interviews, and podcasts (Fig.~\ref{fig:iteration}-(3)). Unlike stripped-down video, these formats preserved continuous narrative engagement while minimizing visual demand. Participants described the experience as both restful and easy to follow: \textit{"I don't want to use my eyes more after staring at a screen all day"} (P1); \textit{"It was relaxing but still interesting"} (P3). This confirmed that audio-centric media effectively mitigates over-immersion and cognitive strain while preserving narrative engagement.

\textbf{Sustaining motivation while reinforcing completion through rhythmic structuring.}
Participants favored breaks that unfolded as an arc: beginning with light movements, escalating to more engaging activities, and concluding with calming exercises for a soft return to work \cite{nguyen2023play} (Fig.~\ref{fig:iteration}-(4)). Seamless guidance proved integral to engagement, making experiences more compelling than passive viewing. When we piloted replacing seamless guidance with non-narrative prompts (Fig.~\ref{fig:iteration}-(5)), participants described them as "forced" and disengaging, reinforcing that guidance must remain media-embedded to feel voluntary. We added a brief ending phase (Fig.~\ref{fig:iteration}-(6)) with lightweight feedback and encouragement, which participants described as helping them transition smoothly back to tasks. This arc, consistent narrative-linked guidance, and dedicated closure constitute our \textbf{Rhythmic Structuring} strategy, sustaining participation while providing clear recovery boundaries.

\begin{figure}[t]
    \centering
    \includegraphics[width=0.5\textwidth]{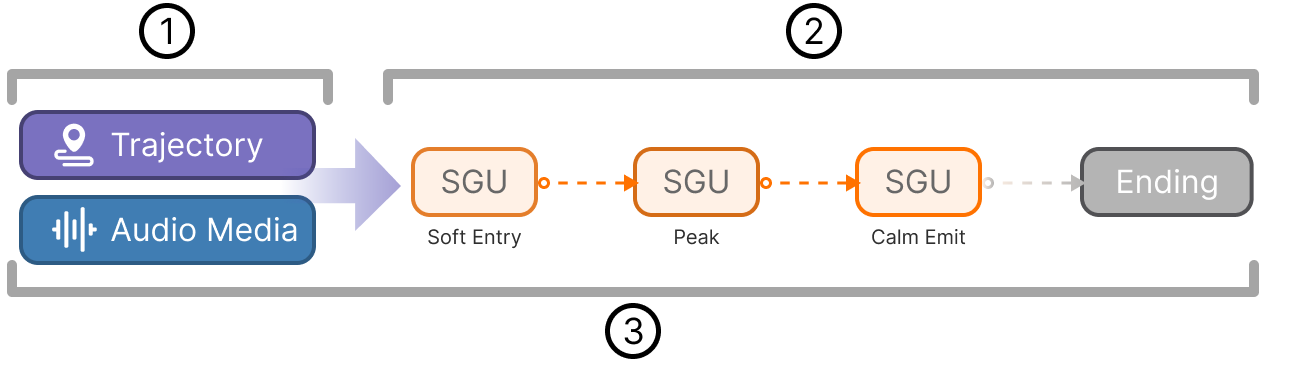}
    \caption{The design pipeline outlines the construction of an AR interactive break experience in three phases: (1) contextual scaffolding, (2) seamless guidance design (see Fig.~\ref{fig:framework}), and (3) holistic coordination and refinement.}
    \Description{Three-phase design pipeline diagram. The flow connects three sequential boxes: (1) Contextual Scaffolding, which establishes the break trajectory and audio media; (2) Seamless Guidance Design, referencing the SGU process; and (3) Holistic Orchestration, which arranges the units into a rise-peak-closure arc.}
    \label{fig:pipeline}
\end{figure}

\added{Together, these strategies determine how individual guidance units are assembled into a complete AR break experience, a process illustrated in Fig.~\ref{fig:pipeline}. Designing the overall session begins with establishing a contextual scaffold that defines both the spatial and temporal structure of the break—namely, a feasible \textbf{Break Trajectory} shaped by workplace layout and social norms, and an \textbf{Audio-Centric Media} substrate that provides continuous narration without increasing visual load. Within this scaffold, \textbf{Seamless Guidance Units (SGUs)} are positioned at suitable points along the trajectory to externalize narrative moments and introduce lightweight restorative actions. Multiple \textbf{Seamless Guidance Units} are then coordinated into a rise–peak–closure arc: early units prompt gentle initiation, mid-session units introduce more engaging embodied movements, and closing units guide users back toward calmness and task readiness. This holistic coordination, refined through iterative adjustment of intensity, modality variation, and spatial placement, balances engagement and recovery across the session.}

\section{Evaluation Study}

This study aims to examine the \added{effectiveness of our media-embedded AR break design, instantiated in the \IBE{} prototype}, by comparing it against three representative break strategies. Our evaluation centers on three \added{core aspects}:

\begin{itemize}
    \item \added{\textbf{Aspect 1 - Seamless transition effectiveness:} To what extent does the \added{media-embedded guidance mechanism} support seamless transitions from media content to break activities?}

    \item \added{\textbf{Aspect 2 - Engagement–restoration balance:} How does the media-embedded AR break experience balance user engagement with restorative break quality? }

    \item \added{\textbf{Aspect 3 - Adaptability and long-term use potential:} How can the \added{media-embedded AR break experience} adapt to varying durations, contexts, and personal preferences to support long-term use?}
\end{itemize}

\added{To investigate these aspects,} we conducted a within-subjects study involving four experimental conditions. These conditions were selected to represent widely observed real-world practices as well as research-supported strategies from prior work \cite{Dan08SuperBreak,Scott17BreakSense,Jiang2024PlayfulAI,Felicia23Mindful}. The design ensured comparability by standardizing the physical activities across conditions.

\subsection{Proof-of-Concept Prototype: \IBE{}}

\begin{figure*}[h]
    \centering
    \includegraphics[width=\textwidth]{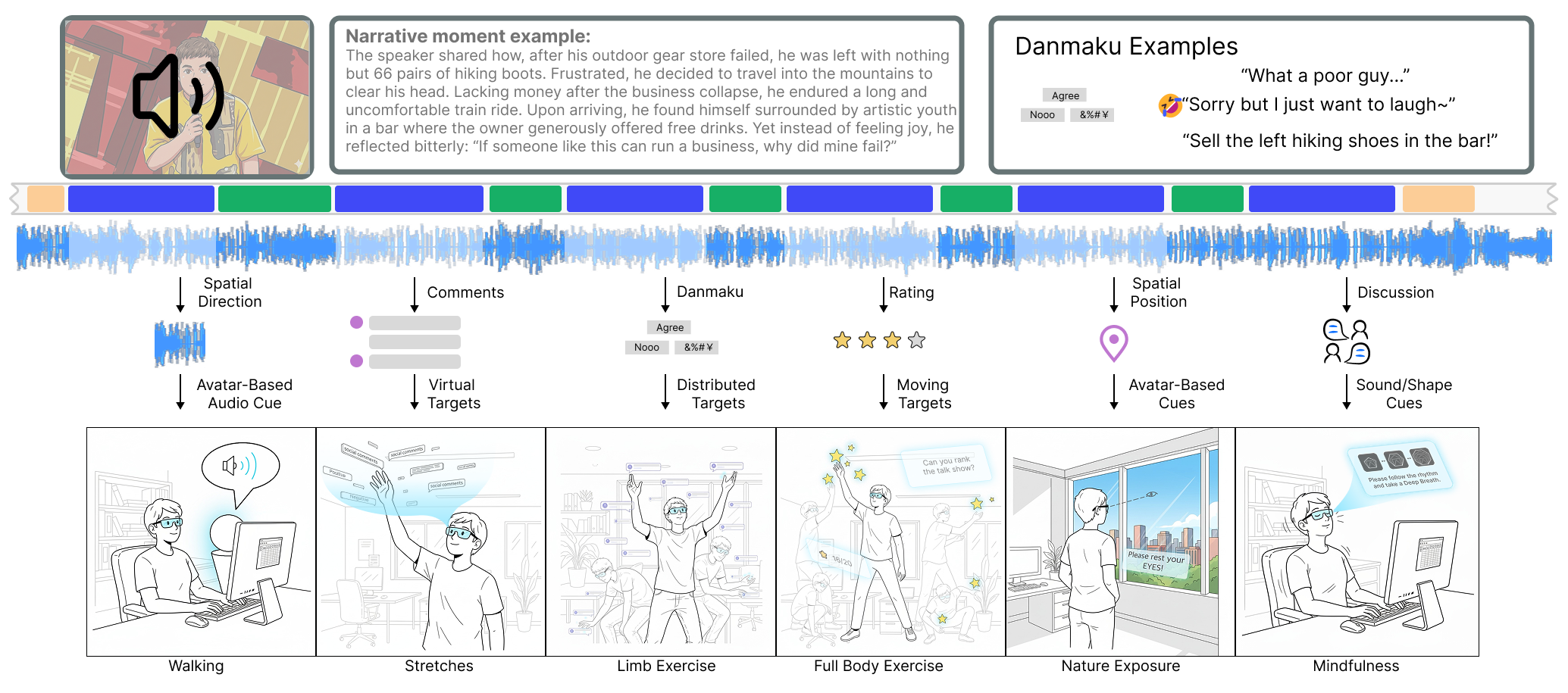}
    \caption{A junior product designer Alex's 15-minute Break with \IBE{}. (1) After a focus block, Alex puts on an AR headset at his desk, where the comedy talk show performer’s voice is spatialized onto an \emph{audio avatar} drifting toward the corridor, prompting her to follow a feasible \emph{Break Trajectory} (desk$\rightarrow$corridor$\rightarrow$window$\rightarrow$desk). (2) A debate segment is externalized into two translucent opinion panels (interactive comments); Alex reaches to “vote,” doubling as a shoulder–neck stretch, confirmed by a soft tick (target visualization + sound cue). (3) After a punchline, drifting \emph{danmaku} bubbles invite a small turn and side-bend to endorse a line (bullet comments, distributed glanceable targets). (4) At the episode’s highlight, a star-rating flourish raises movement amplitude as Alex reaches for moving stars; a bonus clip is unlocked upon completion (rating feature, moving targets). (5) Mid-session, the avatar guides Alex to a window bay for $\sim$30s of distance gaze while the conversation continues (spatial audio prompt). (6) The session closes with audience reflections fading into a short breathing cadence, paired with a summary card and gentle prompt to return to work, completing a five-minute break that sustained engagement while supporting both cognitive and physical recovery.}
    \Description{Storyboard and timeline of a user named Alex experiencing the InteractiveBreak system. A waveform at the top correlates narrative moments with specific activities. Below, six sketches show Alex: (1) Walking following an audio avatar; (2) Stretching to 'vote' on a debate; (3) Leaning to endorse floating comments (danmaku); (4) Reaching for moving stars during a highlight; (5) Gazing out a window during a segue; and (6) Breathing with a visual guide to close the session.}
    \label{fig:Prototype}
\end{figure*}

\added{To evaluate our design framework, we instantiated it as \IBE{}, a deployable proof-of-concept system for everyday office settings (Fig.~\ref{fig:Prototype}).}

\textbf{System setup: contextual scaffolding (Phase 1).}  
The system runs on an AR headset equipped with spatial audio and head/hand tracking. Each session is grounded in a short \emph{Break Trajectory} (desk$\rightarrow$corridor$\rightarrow$window$\rightarrow$desk) that is physically safe and socially unobtrusive for the evaluation environment. The media substrate is an edited, podcast-style comedy talk show: conversational, narrative-rich, and audio-centric, with micro-pauses inserted where actions can be embedded without harming comprehension. 

\textbf{Guidance design: seamless guidance units (Phase 2).}  
Following Sec.~\ref{sec:guidance_design}, candidate \emph{narrative moments} (e.g., topic shifts, pauses, punchlines) were annotated in the episode timeline. Each moment was matched to a locus on the Break Trajectory (e.g., corridor alcoves for small stretches, window areas for gaze). Designers then crafted \emph{Seamless Guidance Units (SGUs)} that externalized the narrative moment into a media element, expressed the paired action via an activity cue, and fused them through semantic, rhythmic, and perceptual alignment. Motivational bindings were added sparingly to sustain participation without diverting attention from recovery.  

The prototype implemented several Seamless Guidance Units (SGUs) to couple narrative moments with lightweight activities:  
\begin{enumerate}
  \item An \emph{audio avatar} carried the host’s voice and drifted forward, transforming continued listening into low-salience locomotion away from and back to the desk (spatial directionality; audio cue; narrative continuation).  
  \item Debate segments were externalized as subtle opinion panels; reaching to “vote” provided shoulder–neck relief while a soft tick confirmed completion (interactive comments; target visualization + brief prompt; light social participation).  
  \item Post-punchline \textit{danmaku} bubbles could be endorsed with a small step, turn, or side-bend, adding light whole-body motion without disrupting comedic timing (bullet comments; distributed targets; playful variation).  
  \item A star-rating mechanic during narrative highlights required quick reaches to moving stars, briefly increasing amplitude and unlocking a bonus clip upon completion (rating feature; moving targets; contingent reveal).  
  \item A mid-session segue redirected the audio avatar toward the window, affording \(20\!-\!30\)s of distance gaze and natural light while the conversation continued (topic-shift cue; spatial audio prompt; emotional attachment).  
  \item The session concluded with audience reflections fading into a short breathing cadence, supported by a low-contrast edge visualization and a concise summary card, providing explicit closure and nudging return to work (audio-comment transformation; audio-visual breathing cue; positive recap).  
\end{enumerate}

\textbf{Session orchestration: rise–peak–closure (Phase 3).}  
The SGUs were sequenced along the Break Trajectory to form a \emph{rise--peak--closure} arc (Sec.~\ref{sec:balance_strategies}). Sessions began with gentle locomotion and simple reaches, escalated into more varied and embodied actions mid-session, and ended with calming exercises that explicitly closed the break at the desk. This orchestration balanced engagement with recovery. The repertoire also supported different durations: a \(\sim\!5\)-minute session included SGUs 1, 2, 4, and 5, while a \(\sim\!15\)-minute session included all six SGUs, with the flexibility to extend the pacing and duration of individual units. In both cases, the continuity of the audio narrative was preserved, ensuring the experience remained coherent and immersive across formats.

\subsection{Experimental Conditions}

We compared four conditions: one video-watching baseline and three activity-based variants. The study adopted two fixed break durations (5 and 15 minutes; see Section~\ref{sec:EvaluationProcedure}). In the video-watching baseline, participants viewed videos of corresponding lengths, while in the activity-based conditions the same activities and timings were used to match these durations. All activity-based conditions followed an identical sequence (or partial) for controlled comparison: walking, upper-body exercise, full-body movement, gazing out the window, and mindfulness (see Figure~\ref{fig:Prototype}).

\begin{itemize}

    \item \textbf{\VW{}.}  
    A passive baseline simulating everyday breaks through media consumption. Participants watched audio-centric comedy talk shows curated from online platforms, matched to break durations.

    \item \textbf{\BR{}.}  
    A reminder-based condition adapted from wellness interventions. Participants received periodic textual and audio reminders via the AR interface, accompanied by a glanceable ambient progress bar~\cite{Lu2021Glanceable}. The design drew from prior work on nudges and ambient feedback~\cite{Yunlong19Progress, Luo2018TimeforBreak}.

     \item \textbf{\INN{}.}  
    An interactive but non-narrative condition, providing the same activity types and guidance cues as other variants but without narrative integration. This baseline isolates the effect of interactive AR guidance alone. For walking, we adapted \textit{BreakSense}~\cite{Scott17BreakSense} with virtual targets and audio feedback. For physical activities, targets inspired by \textit{SuperBreak}~\cite{Dan08SuperBreak} and \textit{BIG-AOME}~\cite{Jiang2024PlayfulAI} prompted reaching and full-body actions. Mindfulness guidance used animated geometric patterns with ambient sound~\cite{Felicia23Mindful}.  

     \item \textbf{\IBE{}.}  
    Our proposed system integrates narrative-driven audio media with interactive activities, enabling smooth transitions from passive viewing to guided movement. Media content was drawn from the same source as \VW{} but reorganized using our authoring framework (Section~\ref{sec:design_framework}), ensuring content consistency while embedding seamless guidance for comparative evaluation.
\end{itemize}

\subsection{Participants and Apparatus}
\textbf{Participants.}
We recruited 16 participants (7 female, 9 male; aged 20–31, M = 24.13, SD = 3.06). All were engaged in screen-intensive knowledge work for more than six hours daily and expressed interest in improving their break habits. On a 7-point Likert scale (where 7 is "very familiar"), participants' self-reported familiarity with AR/MR technology was moderate (M = 4.19, SD = 1.94).

\textbf{Apparatus.}
All four conditions were experienced using the same pair of XREAL Air 2 glasses (52° diagonal field of view) as near-eye AR displays, ensuring consistency across sessions. \IBE{}, \INN{}, and \BR{} conditions were implemented as Android applications developed in Unity 2022.3.6f1, using the XREAL SDK (including built-in hand tracking via the Air 2 Ultra’s cameras), Unity XR Interaction Toolkit, and AR Foundation. All applications were deployed to an XREAL Beam Pro, which handled real-time rendering and interaction. 
The study was conducted in a semi-public, open-plan co-working space on a university campus, comprising individual cubicles, open areas, and connecting hallways to support diverse spatial contexts. All sessions were approved by the workspace manager and present occupants.

\subsection{Procedure}
\label{sec:EvaluationProcedure}
We conducted a within-subjects study where each participant completed four half-day sessions on consecutive workdays, each testing one break condition. Session order was counterbalanced using a Latin Square design. Before beginning, participants provided consent, completed demographics, and received a study overview.

Each session followed a Pomodoro-inspired structure~\cite{cirillo2018pomodoro}: three 25-minute work intervals with 5-minute breaks, a fourth with a 15-minute break, and a final short cycle to assess carryover effects. Participants did not wear AR glasses during work intervals to avoid display occlusion issues with current-generation devices; limiting device use to breaks also mirrored natural digital rest behaviors. 
After each break, participants completed a brief experience questionnaire. Each session concluded with a longer survey on engagement, motivation, usability, and rest quality, followed by a semi-structured interview. After the last session, a final interview was conducted, focusing on the overall experience comparison.

\subsection{Measures and Analysis}
To assess \IBE{}’s seamless transition mechanism (\textbf{Aspect 1}), we collected both subjective perceptions and behavioral responses grounded in our design requirements \cite{zhu2024make, weerasinghe2022arigatoeffectsadaptiveguidance, faas2024choiceconsequencesrestrictingchoices}. Perceived alignment between guidance and media content was assessed only in the \IBE{} condition, while additional 7-point Likert ratings (i.e., transition smoothness, narrative disruption, autonomy, and perceived effort) were gathered across all activity-involved conditions (\IBE{}, \INN{}, \BR{}). Acceptance of guidance was recorded as full, partial, or full rejection. To evaluate the balance between engagement and restorative quality (\textbf{Aspect 2}), we used the short-form User Engagement Scale (UES-SF, 5-point) \cite{o2018practical} and 7-point measures of work-break quality \cite{Epstein2016Taking5}, including overall quality, detachment, relaxation, refreshed state, post-break readiness, and effects on work quality, focus, and productivity. We also assessed pacing perceptions at session stages (entry, middle, return) and the overall sense of completion. To explore adaptability for long-term use (\textbf{Aspect 3}), we measured novelty and sustained interest \cite{shin2019beyond} and obtained final preferences via ranked choice. Semi-structured interviews complemented the surveys, probing experiences of guidance, engagement–restoration balance, contextual fit, reasons for preference, and improvement suggestions. 

Questionnaire data were analyzed using one-way repeated-measures ANOVA with post-hoc pairwise comparisons for significant main effects. \newadded{
Normality assumptions were assessed using the Shapiro-Wilk test with full statistics reported in Appendix~\ref{sec:nomraltest}, and all p-values reported in the Results section are based on these analyses with post-hoc comparisons adjusted using the Bonferroni--Holm correction.} The interview responses analyzing followed the methods in Sec.~\ref{sec:methodoverview} with the codebook provided in Appendix~\ref{sec:appendix_eval}. 

\section{Results}
\label{sec:Results}

Across four half-day sessions, participants consistently rated \IBE{} as the most effective break system compared to the three baselines. It achieved the highest scores across seamlessness (M=6.00/7), overall break quality (M=5.63, SD=0.96), work readiness (M=5.75/7), and engagement (M=4.15/7). In contrast, \VW{}, though entertaining, increased visual fatigue and offered little bodily relief (physical relaxation M=3.25/7). Break Reminders were least effective overall, with the lowest scores for detachment (M=3.88/7) and engagement (M=2.77/7). Interactive Non-Narrative provided moderate physical benefits but was perceived as repetitive and task-like over time. Importantly, 11 out of 16 participants selected \IBE{} as their preferred solution, whereas 9 ranked Break Reminders lowest. These results establish \IBE{} as the most promising candidate for future adoption. 
In the following sections, we detail the mechanisms underlying these outcomes, organized by the three evaluation aspects.

\begin{figure*}[t]
\centering
\includegraphics[width=\textwidth]{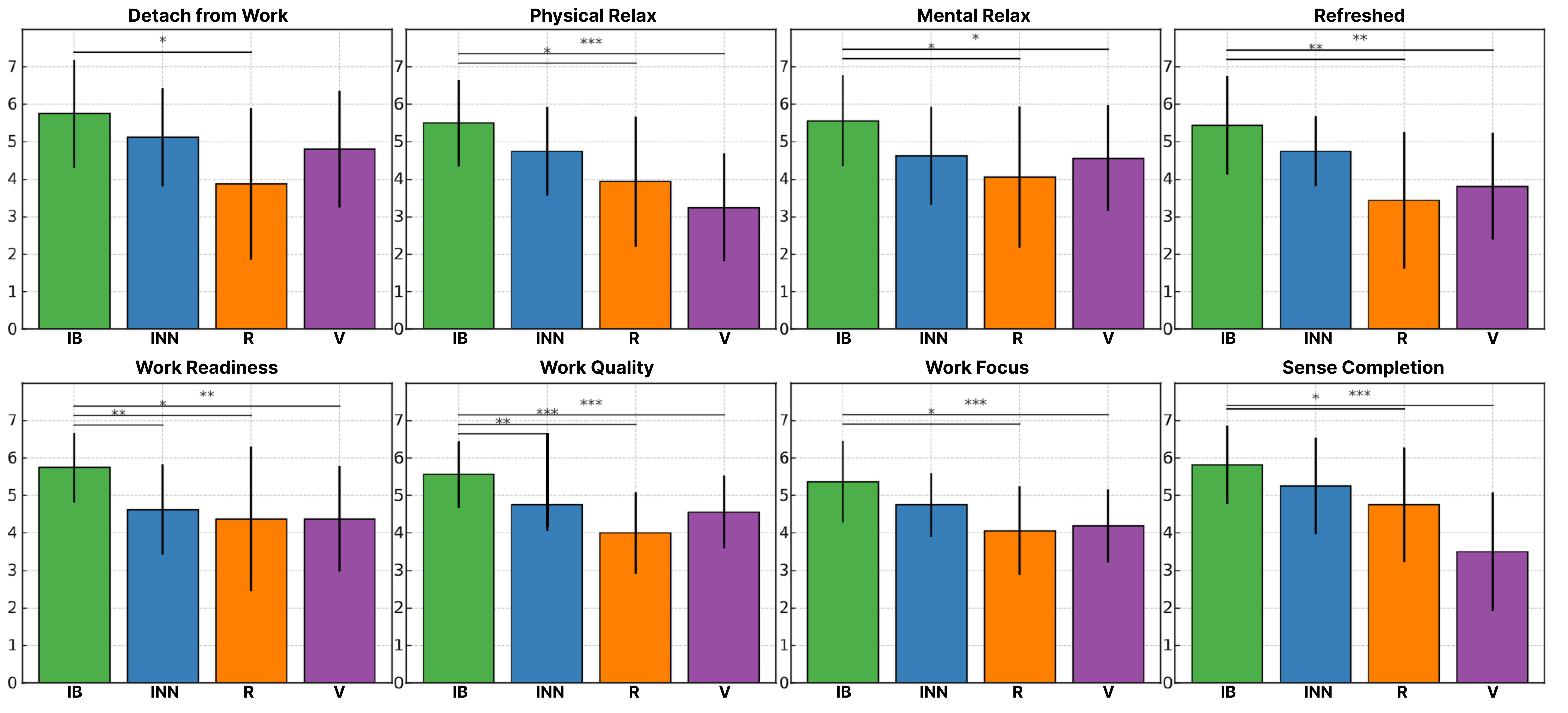}
\caption{Mean likert-7 ratings (±SD) for eight break quality metrics across four conditions (IB = \IBE{}, INN = \INN{}, R = \BR{}, V = \VW{}). Significance markers indicate pairwise comparisons between Interactive Break and the other conditions (*$p < .05$, **$p < .01$, ***$p < .001$). }
\Description{Clustered bar charts comparing four conditions across eight break quality metrics. The conditions are InteractiveBreak (IB), Interactive Non-Narrative (INN), Break Reminder (R), and Video Watching (V). IB scores consistently highest across all metrics: Detach from Work, Physical Relaxation, Mental Relaxation, Refreshed, Work Readiness, Work Quality, Work Focus, and Sense of Completion. Break Reminder (R) generally scores the lowest.}
\label{fig:BreakQuality}
\end{figure*}

\subsection{\added{Aspect 1 - Seamless Transition Effectiveness}}

\textbf{Finding 1: Media-integrated guidance enabled smoother transitions than baseline approaches.} 
Participants rated \IBE{} as offering the most seamless activity transitions. On seamlessness, \IBE{} scored higher (M=6.00, SD=0.97) than both \INN{} (M=4.88, SD=1.02; $p<.01$) and \BR{} (M=3.75, SD=1.91; $p<.01$). Activity introductions were also perceived as less disruptive in \IBE{} (M=5.88, SD=1.26) than in \INN{} (M=4.88, SD=1.15; $p<.05$) and \BR{} (M=4.25, SD=1.57; $p<.01$). Importantly, \IBE{} received high ratings on its media alignment (M=5.56, SD=1.10), suggesting that participants experienced the cues as part of the unfolding narrative rather than as external interruptions. 
Interviews explained this effect by pointing to the integration of guidance with the media flow. Participants emphasized that \IBE{} felt like a continuation of the content rather than a separate prompt. P2 described it as “an extension of the talk show experience,” while P13 noted that “the AR glasses made the activities inseparable from the media.” In contrast, baselines were often characterized as jarring: prompts in \INN{} were predictable and uninteresting (P3, P4, P7, P12), while \BR{} cues were dismissed as intrusive or irrelevant, “just like another notification” (P2, P7, P15).

\textbf{Finding 2: Seamlessness reframed activities as voluntary and approachable.} 
Although the activity set and sequence were identical across conditions, participants rated \IBE{} guidance as more natural and less coercive. On the item “the guidance felt like a suggestion rather than a command,” \IBE{} scored highest (M=5.88, SD=1.15), significantly above \INN{} (M=4.94, SD=1.06; $p<.05$) and \BR{} (M=3.88, SD=1.71; $p<.001$). Physical effort was also judged more reasonable in \IBE{} (M=6.12, SD=0.89) than in both baselines ($p<.05$). These results suggest that integrating cues within the narrative reduced the sense of obligation and reframed the activities as feasible and restorative.  
Qualitative feedback reinforced this interpretation. Participants reported that they did not feel pressured to “do an activity” but were instead guided smoothly through the media experience. As P11 noted, “I never felt forced; \IBE{} just led me there, and afterwards I could feel the benefits.” Similarly, P13 explained that the mechanism encouraged them to begin activities they might normally avoid, while still appreciating the physical benefits once engaged. By contrast, \INN{} prompts were described as mechanical and effort-imposing (P2, P3, P12, P15), and \BR{} cues as abrupt and uninspiring (P1, P4, P9, P10, P15).  

\textbf{Finding 3: Behavioral evidence confirmed higher acceptance of \IBE{}.}
Acceptance rates provided further support: \IBE{} achieved the highest acceptance (M=0.68, SD=0.39), significantly above \BR{} (M=0.07, SD=0.16; $p<.001$) and higher than \INN{} (M=0.51, SD=0.37). Notably, no instances of full rejection occurred in \IBE{}, while both baselines saw occasional or frequent rejections. This behavioral evidence complements the subjective ratings, showing that participants not only perceived \IBE{} as seamless and non-coercive but also acted on it more consistently. 
Together, these findings demonstrate that \IBE{} uniquely preserved immersion by embedding guidance within the media narrative, lowering perceived pressure, and yielding higher behavioral compliance. Whereas \INN{} felt repetitive and task-oriented, and \BR{} intrusive and ignorable, \IBE{} reframed break activities as a natural, voluntary part of the ongoing experience. \added{Some participants disengaged when media or interactions mismatched their preferences, and a few felt visible wearing AR in shared spaces, indicating current social acceptability limits, which are further discussed in Section~\ref{sec:discussion}.}


\subsection{\added{Aspect 2 - Engagement–Restoration Balance}}

\textbf{Finding 1: \IBE{} provides strong restorative quality while also enhancing post-break work outcomes.} 
Across all measures of perceived rest quality, \IBE{} outperformed baselines, especially \BR{}. Participants reported stronger \textbf{Mental Detachment} (M=5.75, SD=1.44) than in \BR{} (M=3.88, SD=2.03; $p=.025$), greater \textbf{Physical Relaxation} (M=5.50, SD=1.15) than in \BR{} (M=3.94, SD=1.73; $p=.010$) and \VW{} (M=3.25, SD=1.44; $p=.001$), and higher \textbf{Mental Relaxation} (M=5.56, SD=1.21) than in \BR{} (M=4.06, SD=1.88; $p=.043$) and \VW{} (M=4.56, SD=1.41; $p=.024$). Participants also felt more \textbf{Refreshed} (M=5.44, SD=1.31) than in \BR{} (M=3.44, SD=1.82; $p=.006$) and \VW{} (M=3.81, SD=1.42; $p=.006$). Like \IBE{}, the \INN{} condition guided participants through varied physical activities, promoting detachment, though average scores were lower and not significantly different on some indicators.
Global measures further favored \IBE{}. \textbf{Work Readiness} was highest (M=5.75, SD=0.93), surpassing \INN{} (M=4.62, SD=1.20; $p=.007$), \BR{} (M=4.38, SD=1.93; $p=.024$), and \VW{} (M=4.38, SD=1.41; $p=.007$). \textbf{Overall Break Quality} was also higher in \IBE{} (M=5.63, SD=0.96) than in \INN{} (M=4.75, SD=0.86; $p=.025$), \BR{} (M=3.88, SD=1.36; $p<.001$), and \VW{} (M=4.25, SD=1.06; $p=.004$). Together, \IBE{} provided both specific rest benefits and stronger overall recovery.

\textit{Impact on Back-at-Work Productivity.}
\IBE{}'s advantages extended to post-break outcomes. \textbf{Work Quality} (M=5.56, SD=0.89) exceeded \INN{} (M=4.75, SD=0.68; $p=.005$), \BR{} (M=4.00, SD=1.10; $p=.001$), and \VW{} (M=4.56, SD=0.96; $p<.001$). \textbf{Focus} improved in \IBE{} (M=5.38, SD=1.09) relative to \BR{} (M=4.06, SD=1.18; $p=.013$) and \VW{} (M=4.19, SD=0.98; $p<.001$), with \INN{} showing only marginal difference ($p=.066$). \textbf{Productivity} was also higher in \IBE{} (M=5.56, SD=0.81) than \INN{} (M=4.75, SD=0.68; $p=.005$), \BR{} (M=4.25, SD=1.29; $p=.006$), and \VW{} (M=4.38, SD=0.89; $p<.001$). Overall, \IBE{} showed clear and significant benefits for most work-related measures.
Interview data clarified these effects. Participants described light activities as fitting their “relaxation needs” (P1), giving “a clear sense of bodily relief” (P8), or releasing tension (P9). They appreciated variety (P7) and the inclusion of practices like stretching or meditation they would not normally attempt (P2). Cognitively, participants highlighted mental relaxation from “listening to the talk show” (P11) and its “joyful, light” content (P4, P13). The audio format reduced eye strain (P5) and helped participants detach from their immediate work environment (P7, P12). Thus, \IBE{} effectively combined light activities with low-visual-demand media, maximizing rest quality and work readiness.

\textit{Baseline conditions.}
For \VW{}, participants agreed it increased eye fatigue and lacked physical activity. While entertaining, it failed to support detachment or bodily recovery: “my body did not relax at all, and I was still straining my eyes on the screen” (P4, P9).
The \BR{} condition was described as weakest: prompts were vague, unengaging, and often ignored. Participants felt they wasted rest time without clear guidance (P9), or even overexerted themselves (P12, P14), returning fatigued.
The \INN{} condition provided bodily relief but did not foster mental detachment. As novelty faded, activities felt like tasks (P2). Some valued its low cognitive demand (P5, P12), but others found it repetitive. While effective for physical relaxation, cognitive effects varied, reflecting differing needs and preferences.

\textbf{Finding 2: \IBE{} sustains calibrated engagement and fosters a stronger sense of completion.} 
On the short-form UES, \IBE{} outperformed \BR{} across all engagement dimensions, with higher focused attention, usability, and aesthetic appeal. Most notably, \IBE{} achieved significantly higher \textit{Reward} scores than \INN{} (M=3.88, SD=0.58; $p=.014$), \BR{} (M=2.90, SD=1.05; $p<.001$), and \VW{} (M=3.50, SD=0.81; $p=.038$), reflecting breaks perceived as intrinsically rewarding rather than task-like. For \textit{Total Engagement}, \IBE{} exceeded \INN{} (M=3.67, SD=0.62; $p=.023$) and \BR{} (M=2.77, SD=0.84; $p<.001$), with no difference from \VW{} (M=3.60, SD=0.75; $p=.078$). Despite lacking visual streams, \IBE{} matched the audiovisual stimulation of video, sustaining strong engagement, see table~\ref{tab:ues_single_col}.

\begin{table}[h]
\centering
\footnotesize 
\setlength{\tabcolsep}{1pt} 

\caption{UES-SF results: \IBE{} compared with other conditions (Single Column Version).}
\Description{Table presenting User Engagement Scale (UES-SF) results. It compares InteractiveBreak against the three baselines across Focused Attention, Perceived Usability, Aesthetic Appeal, Reward, and Total Engagement. InteractiveBreak shows significantly higher scores, particularly in the 'Reward' and 'Total Engagement' categories compared to the Break Reminder and Interactive Non-Narrative conditions.}
\label{tab:ues_single_col}

\begin{tabular}{@{} p{0.30\columnwidth} p{0.52\columnwidth} p{0.18\columnwidth} @{}}
\toprule
\textbf{Metric} & \textbf{Comparison} (M(SD)) & \textbf{Result} ($t, p$) \\
\midrule

\textbf{Focused Attention} \newline \IBE{} \newline \textbf{4.06 (0.76)} 
& \INN{} 3.56 (0.61) \newline \textbf{\BR{} 2.21 (0.93)} \newline \VW{} 3.44 (0.99) 
& 1.62, .126 \newline \textbf{4.96, $<$.001\textsuperscript{***}} \newline 1.78, .095 \\
\midrule

\textbf{Perceived Usability} \newline \IBE{} \newline \textbf{4.42 (0.71)}
& \INN{} 3.92 (0.93) \newline \textbf{\BR{} 3.69 (0.88)} \newline \VW{} 4.13 (0.86) 
& 1.60, .130 \newline \textbf{2.30, .036\textsuperscript{*}} \newline 1.15, .266 \\
\midrule

\textbf{Aesthetic Appeal} \newline \IBE{} \newline \textbf{3.77 (0.59)}
& \INN{} 3.33 (0.78) \newline \textbf{\BR{} 2.27 (1.10)} \newline \VW{} 3.35 (0.76) 
& 2.02, .062 \newline \textbf{4.15, $<$.001\textsuperscript{***}} \newline 1.82, .088 \\
\midrule

\textbf{Reward} \newline \IBE{} \newline \textbf{4.35 (0.63)}
& \textbf{\INN{} 3.88 (0.58)} \newline \textbf{\BR{} 2.90 (1.05)} \newline \textbf{\VW{} 3.50 (0.81)} 
& \textbf{2.79, .014\textsuperscript{*}} \newline \textbf{4.54, $<$.001\textsuperscript{***}} \newline \textbf{2.27, .038\textsuperscript{*}} \\
\midrule

\textbf{Total Engagement} \newline \IBE{} \newline \textbf{4.15 (0.56)}
& \textbf{\INN{} 3.67 (0.62)} \newline \textbf{\BR{} 2.77 (0.84)} \newline \VW{} 3.60 (0.75) 
& \textbf{2.54, .023\textsuperscript{*}} \newline \textbf{4.92, $<$.001\textsuperscript{***}} \newline 1.90, .078 \\
\bottomrule
\end{tabular}
\end{table}

\IBE{} also provided a more balanced and structured experience. Participants rated it significantly more \textbf{Balanced} (M=5.62, SD=0.96) than \BR{} (M=3.12, SD=1.50; $p<.001$) and \VW{} (M=4.56, SD=1.50; $p=.019$). Across phases, \IBE{} excelled: smoother \textbf{Beginnings} (M=5.38, SD=1.50) than \BR{} (M=3.44, SD=1.86; $p=.012$) and \VW{} (M=3.12, SD=1.54; $p=.001$); more engaging \textbf{Middles} (M=6.00, SD=0.89) than \INN{} (M=4.81, SD=1.28; $p=.020$), \BR{} (M=2.50, SD=1.37; $p<.001$), and \VW{} (M=4.69, SD=1.70; $p=.020$); and stronger \textbf{Endings} (M=5.88, SD=1.15) than \INN{} (M=4.88, SD=1.26; $p=.041$), \BR{} (M=3.44, SD=1.67; $p<.001$), and \VW{} (M=2.81, SD=1.47; $p<.001$). A stronger \textbf{Sense of Completion} in \IBE{} (M=5.81, SD=1.06) than in \BR{} (M=4.75, SD=1.71; $p=.039$) and \VW{} (M=3.50, SD=1.54; $p<.001$) was reported, though not significantly higher than \INN{} (M=5.25, SD=1.44; $p=.208$).

Interviews revealed that \textbf{completion stemmed primarily from rest fulfillment}, with structural cues like rhythm and closure serving as secondary enhancers. As P3 explained, “I didn’t feel like I needed more rest—the experience was rich enough that I felt ready to work again.” Others valued closure mechanisms: rotating comments and breathing (P2), smooth rhythm (P4), and calm breathing at the end (P9).
Baselines often failed to meet rest needs, undermining completion. \VW{} was entertaining but caused eye fatigue and lacked bodily recovery, leading participants to “keep watching” rather than feel restored (P1, P3, P5, P6, P9). \BR{} cues were dismissed as insufficient and often induced “rest anxiety” with the countdown (P1, P3, P4, P9). 

\INN{} supported physical relief but not cognitive detachment; over time, activities felt like “tasks” (P3, P4, P13, P15).
Overall, \IBE{} balanced sustained engagement with fulfillment of rest needs, ensuring participants felt restored and ready to return. By regulating intensity, embedding light physical activities, and structuring beginnings and endings, it avoided over-immersion while fostering a strong sense of completion.
At the same time, individual differences shaped experiences. Some found talk-show content cognitively demanding (P1, P6, P11, P14, P15), preferring lighter formats. Others valued the simplicity of \INN{} for “mental zoning out,” despite reduced novelty. These differences underscore the need for adaptive break design, a point elaborated in later sections. 

\subsection{\added{Aspect 3 - Adaptability and Long-Term Use Potential}}
\textbf{Finding 1: \IBE{} shows potential for long-term adoption through freshness and adaptability.} 
In final preference ratings, 11 of 16 participants chose \IBE{} as their most preferred method for long-term use, while \BR{} was least preferred (ranked lowest by 9 participants). On \textbf{novelty and sustained interest}, \IBE{} scored highest (M=5.62, SD=1.09), significantly above \INN{} (M=4.00, SD=1.41; $p=.010$), \BR{} (M=2.25, SD=1.65; $p<.001$), and \VW{} (M=4.12, SD=1.67; $p=.006$). Embedding diverse media with integrated activities allowed \IBE{} to sustain novelty across cycles.

\begin{figure}[t]
\centering
\includegraphics[width=0.5\textwidth]{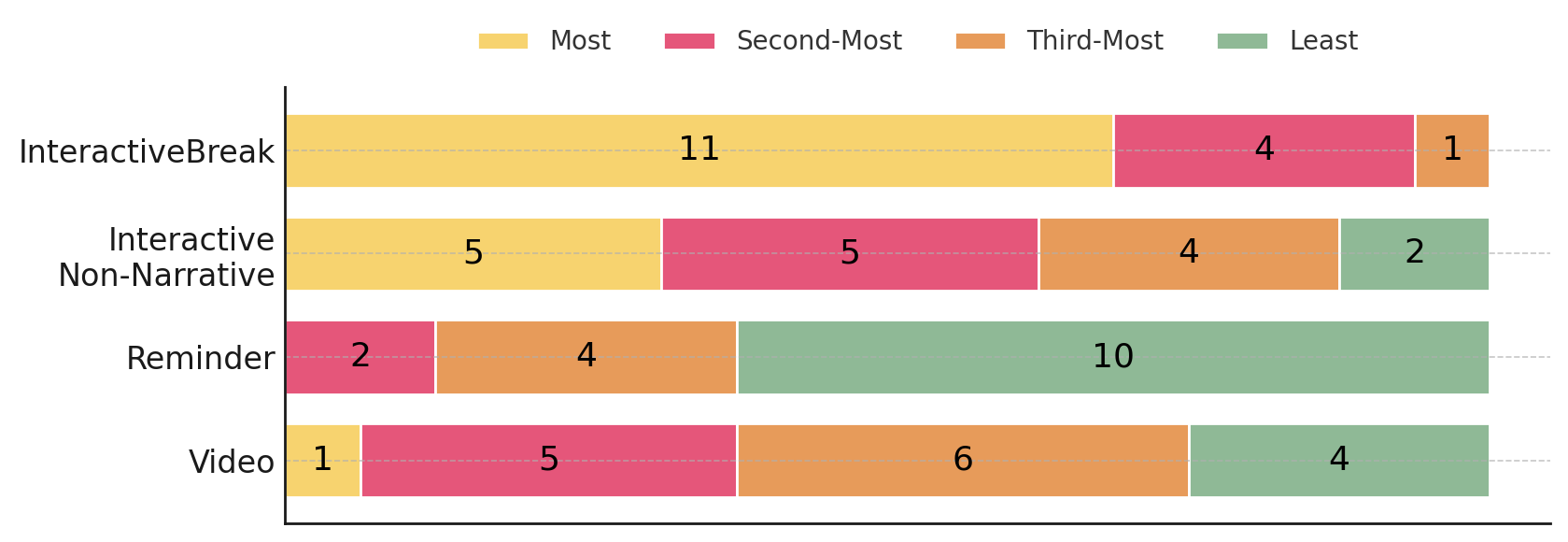}
\caption{Participants' preference rankings for long-term use across the four conditions. InteractiveBreak was ranked as the most preferred option by 11 out of 16 participants.}
\Description{Stacked bar chart of participant preference rankings for long-term use. The chart displays preferences for four conditions: InteractiveBreak, Interactive Non-Narrative, Reminder, and Video. InteractiveBreak is ranked 'Most Preferred' by 11 out of 16 participants. The Reminder condition is ranked 'Least Preferred' by 10 out of 16 participants.}
\label{fig:preference}
\end{figure}

Interviews explained this durability. Participants noted that novelty was preserved because media content varied each session (P3, P10, P11, P13), with repeated activities less noticeable when focus was on the changing narrative (P9). Some suggested refinements such as greater path diversity or more activity types (P2, P7, P9). Many emphasized adaptability: concise interactions suited short breaks (P3, P4, P10), while longer breaks could incorporate richer paths, broader activity ranges, and pacing adjustments (P7, P9). Overall, participants saw \IBE{} as broadly applicable beyond talk shows, adaptable to different media sources, durations, and preferences.

Baseline conditions highlighted these contrasts. \VW{} maintained novelty through varied content but induced visual fatigue, prolonged sitting, and risked over-immersion, making it harder to resume work (P3, P4, P9). \BR{} was seen as monotonous, ignorable, and ineffective, often producing “time anxiety” from the countdown without delivering recovery (P1, P4, P5, P9). \INN{} initially engaged users with light movement but quickly became “predictable and task-like” (P2, P3, P14), with motivation dropping over repeated use (P9, P15). While some game enthusiasts appreciated its simplicity (P6, P11), most found it limited in variety and unsustainable over longer sessions.
Taken together, these findings show why \IBE{} demonstrated the strongest long-term potential. Unlike \VW{}, it avoided over-immersion while sustaining break quality; unlike \BR{}, it actively motivated participation; and unlike \INN{}, it maintained freshness through dynamic media integration rather than fixed routines. By combining media-driven novelty with embedded bodily activity, \IBE{} offered a versatile and sustainable pathway for high-quality breaks.

\section{Discussion}
\label{sec:discussion}
\added{Our findings reveal how media-embedded guidance can effectively support transitions into restorative break activities when three conditions are met: the guidance must feel seamlessly integrated with the ongoing media experience, the break must satisfy both cognitive and physical rest needs, and the activities must be feasible and socially acceptable in the current setting. Synthesizing these insights, we discuss the design principles that govern such experiences. We also explore how personalization and context-aware adaptation can sustain these restorative breaks over time, and how the underlying mechanism applies across scenarios, behaviors, and device ecologies.}

\subsection{Design Insights for Creating Effective Media-Embedded Break Experiences}
\label{sec:design-insights}

\added{Across our iterative prototyping and evaluation, participants’ responses surfaced a set of conditions that determined whether guidance felt seamless, whether the break felt restorative, and whether activities were socially and physically acceptable. We synthesize these observations into design insights that offer practical guidance for future systems. }

\subsubsection{\textbf{Seamless guidance requires media-led immersion and actionable activities}}

\added{Participants’ experiences indicate that seamlessness in \IBE{} arose not from making guidance subtler, but from letting the media itself lead the transition and pairing it with actions that were immediately doable. Our evaluation helps contextualize how this mechanism differs from the baseline approaches. Reminder-based prompts~\cite{Luo2018TimeforBreak,Yunlong19Progress} often lacked clear actionable direction, offered limited motivational pull, and were perceived as interruptions that participants either ignored or replaced with their own digital break routines. Interactive non-narrative cues~\cite{Scott17BreakSense,Dan08SuperBreak} drew initial interest but soon felt repetitive, and the activities increasingly resembled assigned tasks rather than self-chosen rest. In contrast, participants reported that our media-embedded approach clarified what to do through in-context cues, sustained motivation by pairing small, doable actions with ongoing media engagement, and reduced disruption by letting guidance emerge naturally from the content. They also noted that varying media content across sessions helped maintain freshness, and that personalization of media sources in future iterations could further support long-term appeal. This mechanism reflects a micro-task principle, in which a larger goal is decomposed into short, immediately actionable steps that lower initiation friction~\cite{cheng2015break}. Applied to restorative breaks, embedding small, achievable actions directly into the unfolding media flow enabled participants to begin activities without perceiving them as separate tasks. Rather than relying on guilt, restraint, or self-control strategies~\cite{kim2019goalkeeper,hiniker2016mytime,lu2024interactout}, this approach motivated participation by preserving immersion and making healthy actions feel like a natural part of the experience.}

\subsubsection{\textbf{Restorative completion requires aligning pacing and activities with rest needs}}
\added{Our findings show that a break feels “complete” only when its design simultaneously regulates engagement through pacing and caters to both cognitive and physical rest needs—two aspects that existing break practices and interventions often fail to address together. Common habits such as watching videos during breaks~\cite{Duarte2014Web,Epstein2016Taking5} offer momentary cognitive lightness but also introduce visual strain, prolonged sitting, and over-immersion that stretches breaks beyond their intended duration. Reminder-based interventions~\cite{Luo2018TimeforBreak,Yunlong19Progress} prompt timely breaks but were not designed to support deeper detachment, leading some participants to resume habitual media afterward. Lightweight interactive systems~\cite{Scott17BreakSense,Dan08SuperBreak} encourage brief physical activation, yet their simple, low-variation interactions offer limited support for sustained mental disengagement. Our findings complement these approaches by showing that restoration is strengthened when pacing and activities are intentionally aligned: low-visual-load media facilitate cognitive detachment, light micro-activities support bodily release and gradual unwinding, and a calming closure helps users settle before returning to work. Participants repeatedly described this combination as offering “a clear sense of both mental and bodily relief” and being “rich enough that I felt ready to work again.” This aligns with the Effort–Recovery Model~\cite{meijman2013psychological}, which posits that different work demands deplete different resources; our results extend this principle to short breaks by showing that recovery depends on sequencing media–activity elements to target whichever resources (e.g., eye fatigue, cognitive load, bodily tension) are most taxed in the moment. Accordingly, pacing should function not as an abstract curve but as a mechanism for orchestrating these elements in the right order. A remaining challenge is that rest needs vary widely across individuals and contexts, underscoring the importance of adapting both media and activity composition for personalized restorative completeness.}

\subsection{Personalization and Context-Aware Adaptation to Sustain Seamless and Restorative Break Experiences}

\added{To sustain the two mechanisms identified above, seamless media-led guidance and restorative completeness, future systems must adapt to 1) users’ evolving rest needs, 2) personal preferences, and 3) situational constraints. For seamless guidance, the experience must first keep users comfortably immersed; this requires personalizing both media content and media format so that the ongoing narrative aligns with what users naturally enjoy and can cognitively absorb. 
Moreover, the system should adapt to users’ preferred motivational styles, whether driven by narrative suspense \cite{cheng2023designing}, emotional attachment \cite{murnane2020designing,zhao2024grow} or social connections \cite{Scott17BreakSense,Yuki2023AcousticAR2}, and adjusting the modality, granularity, and spatial–visual load of cues so that transitions feel like an organic continuation of the media rather than an external command~\cite{cai2024pandalens}. 
Beyond seamlessness, restorative completeness depends on matching the break’s composition to users’ shifting recovery demands \cite{meijman2013psychological}: some moments call for low-visual-load audio or simple soundscapes to support cognitive lightness, whereas others benefit from richer imagery to promote mental defocusing. Correspondingly, activity type, intensity, and overall pacing must flex to both cognitive and bodily states, ranging from light mental drift to accumulated muscular fatigue. Participants’ accounts reflected this diversity: some preferred lighter, low-demand audio for unwinding (P15), others noted that additional sensory cues helped them transition out of work mode (P7), while some felt the existing audio format already matched their needs well (P4). Differences in bodily and cognitive fatigue further shaped what felt restorative in the moment --- for example, alternating between shoulder relief, waist stretches, or brief mental rest (P1), or finding that mismatched pacing or reach distances reduced comfort and immersion (P2, P14).}

\added{Real workplaces further impose spatial and social constraints that can disrupt otherwise effective guidance; systems therefore need to infer physical layout, social exposure, and momentary comfort, scaling activities from full-body motions to subtle gestures \cite{jindu_handows} or purely attentional shifts as needed. Throughout the experience, the system must also adjust how much agency it offers, respecting users’ preferences for directive versus flexible guidance~\cite{zhang2025neurosync}, and shifting the available options when the environment makes certain actions less feasible or socially comfortable~\cite{Fang2025SocialSimulation}. This sensitivity is essential, as some participants paused activities when others passed nearby because it made them feel watched (P10), indicating that systems must modulate visibility, intensity, and available alternatives to maintain social comfort in dynamic workplaces~\cite{cai2023paraglassmenu}. Together, these forms of adaptation, media fit, motivational alignment, recovery-oriented activity calibration, socio-physical sensitivity, and agency flexibility, form an adaptive design layer that supports seamless and restorative break experiences across diverse contexts. }

\subsection{Design Generalizability Across Scenarios, Behaviors, and Devices}

\added{The core mechanism of \IBE{}, embedding guidance into engaging media and orchestrating transitions into effortful actions, generalizes across scenarios, behaviors, and devices. \textbf{Scenario-wise}, while instantiated for desk work, \IBE{} applies wherever media consumption co-occurs with restorative opportunities: static indoor contexts (meeting pauses, study breaks) support subtle posture resets; domestic situations (cooking, household tasks) allow background media with micro-movements; shared breaks (group coffee breaks) enable lightweight actions if socially acceptable; transitional contexts (commuting, walking) support gentle movements or breathing when spatially appropriate. \textbf{Behavior-wise}, the challenge of shifting from comfortable engagement into effortful action extends beyond breaks. When cues align with narrative or rhythmic moments, they scaffold attention so actions feel natural, supporting micro-learning~\cite{Cai2025Aiget} during leisure, light rehabilitation during listening, or interrupting habitual checking~\cite{hiniker2016mytime} with breathing cues. Rise–peak–closure pacing helps users feel completion while avoiding over-immersion, relevant for learning sessions, mindfulness routines~\cite{Felicia23Mindful}, or reducing compulsive scrolling~\cite{Lee25PurposeMode}. \textbf{Device-wise}, the mechanism requires only media holding attention and interaction channels delivering aligned cues: smartphones embed guidance into video, music, or podcasts with visual/haptic cues and IMU sensing~\cite{xia2022virtual}; smartwatches and earables provide discreet pathways through wrist sensors and audio~\cite{Xia2024AudioMove}; TVs, ambient displays and smart speakers combine playback with activity detection~\cite{ball2021scaling}; AR glasses reduce visual burden while offering multimodal richness and spatial feedback, advantages harder to achieve with pure audio or haptics. Future work should explore how platforms distribute media presentation, cue delivery, and sensing while preserving these principles.}

\subsection{Limitations}
While our findings highlight \IBE{}’s potential for embedding restorative activities into everyday media, several limitations frame the scope of this work. First, the evaluation was constrained by AR hardware. Even with a lightweight headset (XREAL Air2 Ultra), the limited field of view, color distortion, and added weight may have influenced comfort and guidance. To control variance, all conditions used the same device, but restricting use to scheduled breaks, rather than “always-on” wear, reduced ecological validity relative to our envisioned deployment. Second, the experimental design prioritized comparability over realism. The fixed Pomodoro-inspired schedule (25/5/15 minutes) enabled condition contrasts but does not reflect the irregularity of real-world breaks \cite{Epstein2016Taking5}. Likewise, the four-day study probed feasibility through repeated exposure but cannot address long-term adherence or novelty decay, underscoring the need for extended in-situ deployments. Third, we standardized media with the same edited talk-show content. While this ensured consistency, it inevitably mismatched some preferences. Participants anticipated benefits across other genres, but generality remains untested. Overall, these limitations reflect trade-offs for experimental control. Our study serves as a proof-of-concept, laying the groundwork for future research to extend the mechanism across richer media, varied contexts, and long-term workplace use.

\section{Conclusion}
We explored how wearable AR can transform breaks for young knowledge workers from passive media use into restorative, engaging experiences. Through a co-design process, we developed a design framework and \IBE{} emphasizing two key components: (1) dual alignment: linking media with lightweight activities; and (2) calibrated engagement: using audio-driven arcs that sustain motivation while preserving rest.
A within-subjects study (N=16) showed that embedding guidance within media improves transitions into activity, enhances break quality, readiness to return, and long-term preference. 
More broadly, our contributions highlight AR's potential to borrow media engagement to support low-effort health behaviors in everyday contexts. The \textit{Seamless Guidance Unit} offers a reusable pattern for coupling content with context-appropriate actions, suggesting scalable interventions for open workspaces. We further highlight the key to designing for enough engagement rather than full immersion, by keeping users in the narrative, inviting restorative acts, and closing decisively.

\begin{acks}
The authors sincerely thank all participants involved in the various stages of our studies for generously sharing their time, experiences, and insights. We are also deeply grateful to the anonymous reviewers and community members for their constructive feedback and support. This work was partially supported by the Research Grants Council (RGC) of Hong Kong under the RGC GRF grant 16214623.
\end{acks}

\bibliographystyle{ACM-Reference-Format}
\bibliography{References}

\clearpage   
\onecolumn   
\appendix    

\section{Co-design Process}
\label{sec:appendix_codesign}

\subsection{Design Probe Study}

\begin{table}[h]\small
    \caption{Qualitative codebook showing how participants perceived (1) the effectiveness of media-driven activity guidance and (2) the restorative qualities of the overall break experience. Subthemes were derived through axial coding, with representative quotes illustrating each pattern.}
    \Description{Qualitative codebook from the design probe study. It lists two main themes: Activity-Guidance Mechanism and Overall Break Experience. Subthemes describe natural guidance, embodied actions, mental lightness, and the risk of over-immersion, supported by participant quotes.}
    \label{tab:probe_codebook}
        \begin{tabular}{p{0.17\textwidth} p{0.32\textwidth} p{0.435\textwidth}}
        \toprule
        \textbf{\added{Theme}} & \textbf{\added{Subtheme (Axial Coding)}} & \textbf{\added{Example Quote}} \\
        \midrule
        
        \multirow{5}{=}{\textbf{\parbox{2.3cm}{\added{Activity-Guidance}\\\added{Mechanism}}}}
        & \added{Guidance feels natural when emerging organically from the narrative.}
        & \textit{\added{"The transition felt so smooth, like the story itself continued into my room and carried me along."}} \added{(P6)} \\
        \cmidrule(lr){2-3}
        
        & \added{Narrative-driven prompts increase willingness to follow.}
        & \textit{\added{"If the prompt comes from the storyline, like a reveal or a character action, I naturally want to follow it."}} \added{(P5)} \\
        \cmidrule(lr){2-3}
        
        & \added{Embodied actions motivate participation when intuitively tied to media events.}
        & \textit{\added{"When the character hit, I hit; it matched the scene and felt fun without needing extra thought."}} \added{(P3)} \\
        \cmidrule(lr){2-3}
        
        & \added{Actions must stay lightweight, simple, and feasible within the workspace.}
        & \textit{\added{"Movements should be simple and doable instantly, nothing that requires precision or thinking."}} \added{(P2)} \\
        \cmidrule(lr){2-3}
        
        & \added{Context misalignment (physical or social) breaks immersion.}
        & \textit{\added{"Some actions just don't suit where I'm standing; if it feels mismatched with my environment, immersion breaks."}} \added{(P4)} \\
        \midrule
        
        \multirow{5}{=}{\textbf{\parbox{2.3cm}{\added{Overall Break}\\\added{Experience}}}}
        & \added{Mental lightness supports restorative detachment from work.}
        & \textit{\added{"It let me detach from work without needing to think—the experience felt mentally light."}} \added{(P7)} \\
        \cmidrule(lr){2-3}
        
        & \added{Light physical activation aids recovery without feeling effortful.}
        & \textit{\added{"I moved without realizing it; it felt refreshing rather than like deliberate exercise."}} \added{(P1)} \\
        \cmidrule(lr){2-3}
        
        & \added{Novelty and variety sustain engagement across sessions.}
        & \textit{\added{"Every round felt a little different, new moments kept me curious and willing to follow again."}} \added{(P8)} \\
        \cmidrule(lr){2-3}
        
        & \added{Over-immersion may blur break boundaries and hinder return to work.}
        & \textit{\added{"If it gets too immersive, I wouldn't want the break to end; it becomes harder to switch back to work."}} \added{(P4)} \\
        \cmidrule(lr){2-3}
        
        & \added{Continuous visual attention can limit opportunities for eye rest.}
        & \textit{\added{"Because I kept watching the screen, my eyes didn't really rest; audio moments would help me relax more."}} \added{(P5)} \\
        \bottomrule
    \end{tabular}\vspace*{-12pt}
\end{table}

\section{Evaluation Study}
\label{sec:appendix_eval}

\subsection{Participant Information}

\begin{table}[h]
    \caption{Participant demographics and work characteristics. All participants engaged primarily in cognitively intensive, screen-based tasks consistent with knowledge work.}
    \Description{Participant demographics table. It lists 16 participants (P1-P16) with their Gender, Age (ranging 20-31), Knowledge Work Domain (Analytical, Design, or Research), Self-reported AR Familiarity (rated 1-7), and Average Daily Screen Time (mostly greater than 8 hours).}
    \label{tab:participants}
        \begin{tabular}{lccccc}
        \toprule
        \added{\textbf{ID}} & \added{\textbf{Gender}} & \added{\textbf{Age}} & \added{\textbf{Knowledge Work Domain}} & \added{\textbf{AR/MR Familiarity (1–7)}} & \added{\textbf{Avg. Screen Time}} \\
        \midrule
        \added{P1}  & \added{M} & \added{21} & \added{Analytical \& Computational Work} & \added{4} & \added{6–8h} \\
        \added{P2}  & \added{M} & \added{20} & \added{Analytical \& Computational Work} & \added{6} & \added{6–8h} \\
        \added{P3}  & \added{F} & \added{24} & \added{Design \& Creative Work} & \added{3} & \added{>8h} \\
        \added{P4}  & \added{F} & \added{26} & \added{Research \& Academic Work} & \added{3} & \added{>8h} \\
        \added{P5}  & \added{F} & \added{24} & \added{Design \& Creative Work} & \added{5} & \added{>8h} \\
        \added{P6}  & \added{M} & \added{23} & \added{Analytical \& Computational Work} & \added{2} & \added{>8h} \\
        \added{P7}  & \added{M} & \added{26} & \added{Analytical \& Computational Work} & \added{7} & \added{6–8h} \\
        \added{P8}  & \added{M} & \added{21} & \added{Design \& Creative Work} & \added{2} & \added{>8h} \\
        \added{P9}  & \added{F} & \added{26} & \added{Design \& Creative Work} & \added{1} & \added{>8h} \\
        \added{P10} & \added{M} & \added{31} & \added{Analytical \& Computational Work} & \added{7} & \added{>8h} \\
        \added{P11} & \added{F} & \added{24} & \added{Design \& Creative Work} & \added{7} & \added{>8h} \\
        \added{P12} & \added{F} & \added{21} & \added{Artistic \& Media Production} & \added{5} & \added{6–8h} \\
        \added{P13} & \added{M} & \added{25} & \added{Analytical \& Computational Work} & \added{4} & \added{6–8h} \\
        \added{P14} & \added{M} & \added{25} & \added{Analytical \& Computational Work} & \added{2} & \added{>8h} \\
        \added{P15} & \added{F} & \added{22} & \added{Artistic \& Media Production} & \added{5} & \added{>8h} \\
        \added{P16} & \added{M} & \added{27} & \added{Analytical \& Computational Work} & \added{4} & \added{>8h} \\
        \bottomrule
    \end{tabular}\vspace*{-12pt}
\end{table}

\subsection{Shapiro-Wilk Test}
\label{sec:nomraltest}

\begin{table}[h]
    \caption{\newadded{Shapiro-Wilk Test Results (W statistic and p-value)}}
    \Description{Statistical table of Shapiro-Wilk Test results. It lists the W statistic and p-value for all measured metrics (such as Seamlessness, Engagement, and Break Quality) across the four experimental conditions, used to verify the normality assumption for data analysis.}
    \resizebox{\textwidth}{!}{%
        \begin{tabular}{l|c|c|c|c}
            \hline
            \textbf{\newadded{Metric}} & \textbf{\newadded{Basic Interaction}} & \textbf{\newadded{Interactive Break}} & \textbf{\newadded{Reminder}} & \textbf{\newadded{Video Watching}} \\
            \hline
            \newadded{Seamlessness} & \newadded{W=0.862, p=0.021} & \newadded{W=0.800, p=0.003} & \newadded{W=0.941, p=0.355} & \newadded{N/A} \\
            \newadded{Non-disruption} & \newadded{W=0.873, p=0.030} & \newadded{W=0.751, p=0.001} & \newadded{W=0.860, p=0.020} & \newadded{N/A} \\
            \newadded{Suggestions rather than commands} & \newadded{W=0.899, p=0.076} & \newadded{W=0.829, p=0.007} & \newadded{W=0.911, p=0.121} & \newadded{N/A} \\
            \newadded{Reasonable Physical Efforts} & \newadded{W=0.827, p=0.006} & \newadded{W=0.827, p=0.006} & \newadded{W=0.889, p=0.055} & \newadded{N/A} \\
            \newadded{Acceptance} & \newadded{W=0.867, p=0.025} & \newadded{W=0.784, p=0.002} & \newadded{W=0.539, p<0.001} & \newadded{N/A} \\
            \newadded{Focused Attention} & \newadded{W=0.936, p=0.304} & \newadded{W=0.920, p=0.170} & \newadded{W=0.924, p=0.194} & \newadded{W=0.863, p=0.021} \\
            \newadded{Perceived Usability} & \newadded{W=0.917, p=0.152} & \newadded{W=0.806, p=0.003} & \newadded{W=0.949, p=0.469} & \newadded{W=0.843, p=0.011} \\
            \newadded{Aesthetic Appeal} & \newadded{W=0.929, p=0.233} & \newadded{W=0.927, p=0.218} & \newadded{W=0.925, p=0.201} & \newadded{W=0.916, p=0.146} \\
            \newadded{Reward} & \newadded{W=0.950, p=0.490} & \newadded{W=0.864, p=0.022} & \newadded{W=0.918, p=0.158} & \newadded{W=0.953, p=0.542} \\
            \newadded{Total Engagement} & \newadded{W=0.931, p=0.255} & \newadded{W=0.938, p=0.325} & \newadded{W=0.936, p=0.305} & \newadded{W=0.918, p=0.157} \\
            \newadded{Detach from Work} & \newadded{W=0.869, p=0.027} & \newadded{W=0.769, p=0.001} & \newadded{W=0.922, p=0.180} & \newadded{W=0.813, p=0.004} \\
            \newadded{Physical Relax} & \newadded{W=0.821, p=0.005} & \newadded{W=0.812, p=0.004} & \newadded{W=0.874, p=0.031} & \newadded{W=0.876, p=0.034} \\
            \newadded{Mental Relax} & \newadded{W=0.831, p=0.007} & \newadded{W=0.836, p=0.009} & \newadded{W=0.865, p=0.023} & \newadded{W=0.946, p=0.429} \\
            \newadded{Refreshed} & \newadded{W=0.862, p=0.020} & \newadded{W=0.862, p=0.020} & \newadded{W=0.931, p=0.253} & \newadded{W=0.794, p=0.002} \\
            \newadded{Work Readiness} & \newadded{W=0.922, p=0.180} & \newadded{W=0.862, p=0.020} & \newadded{W=0.920, p=0.171} & \newadded{W=0.883, p=0.044} \\
            \newadded{Overall Break Quality} & \newadded{W=0.759, p=0.001} & \newadded{W=0.866, p=0.024} & \newadded{W=0.901, p=0.083} & \newadded{W=0.931, p=0.255} \\
            \newadded{Work Quality} & \newadded{W=0.796, p=0.002} & \newadded{W=0.888, p=0.051} & \newadded{W=0.822, p=0.005} & \newadded{W=0.853, p=0.015} \\
            \newadded{Work Focus} & \newadded{W=0.884, p=0.044} & \newadded{W=0.833, p=0.008} & \newadded{W=0.921, p=0.178} & \newadded{W=0.856, p=0.017} \\
            \newadded{Work Productivity} & \newadded{W=0.796, p=0.002} & \newadded{W=0.827, p=0.006} & \newadded{W=0.944, p=0.394} & \newadded{W=0.870, p=0.027} \\
            \newadded{Beginning} & \newadded{W=0.891, p=0.058} & \newadded{W=0.829, p=0.007} & \newadded{W=0.939, p=0.336} & \newadded{W=0.919, p=0.163} \\
            \newadded{Middle} & \newadded{W=0.845, p=0.011} & \newadded{W=0.776, p=0.001} & \newadded{W=0.876, p=0.033} & \newadded{W=0.896, p=0.069} \\
            \newadded{Ending} & \newadded{W=0.908, p=0.108} & \newadded{W=0.829, p=0.007} & \newadded{W=0.897, p=0.071} & \newadded{W=0.918, p=0.157} \\
            \newadded{Sense of Completion} & \newadded{W=0.892, p=0.060} & \newadded{W=0.840, p=0.010} & \newadded{W=0.783, p=0.002} & \newadded{W=0.919, p=0.160} \\
            \newadded{Novelty and sustained interest} & \newadded{W=0.852, p=0.015} & \newadded{W=0.880, p=0.039} & \newadded{W=0.779, p=0.001} & \newadded{W=0.829, p=0.007} \\
            \hline
        \end{tabular}%
    }
  
    \label{tab:shapiro_wilk}
\end{table}

\subsection{Codebook}

\begin{table*}[t]
\small
\setlength{\tabcolsep}{6pt}
\caption{Qualitative codebook summarizing how participants perceived (1) the seamlessness of media-driven activity guidance, (2) the balance between engagement and restoration, and (3) the long-term adaptability of InteractiveBreak. Subthemes were derived through axial coding, with representative quotes illustrating each perceptual pattern.}
\Description{Qualitative codebook from the evaluation study. It details three major themes: Seamless Guidance, Engagement-Restoration Balance, and Sustained Novelty and Adaptability. Each theme includes subthemes derived from axial coding and representative quotes from participants regarding their experience with the final prototype.}
\label{tab:full_codebook}

\begin{tabularx}{\textwidth}{p{0.18\textwidth} p{0.30\textwidth} X}
\toprule
\textbf{Theme} & \textbf{Subtheme (Axial Coding)} & \textbf{Representative Quote} \\
\midrule

\multirow{5}{=}{\textbf{Seamless Guidance}}
& Narrative-anchored transitions feel natural.
& \textit{``The shift into the activity felt like part of the story—almost as if the media itself led me into the movement.''} (P8) \\
\cline{2-3}

& Media--action alignment increases willingness to follow.
& \textit{``When the prompts came from the storyline, I naturally wanted to follow; it didn't feel like an external instruction.''} (P5) \\
\cline{2-3}

& Clear, actionable prompts support fluid transitions.
& \textit{``The cues were simple and doable instantly, so I never felt stuck or unsure of what to do.''} (P2) \\
\cline{2-3}

& Guidance is perceived as suggestion rather than command.
& \textit{``It felt like a gentle suggestion—something I could choose to join—rather than a task I had to complete.''} (P11) \\
\cline{2-3}

& Contextual fit (physical or social) shapes seamlessness.
& \textit{``If the action doesn't match the space I'm in, immersion breaks immediately; I won't follow something that feels out of place.''} (P4) \\

\midrule

\multirow{6}{=}{\textbf{Engagement--Restoration Balance}}
& Cognitive detachment and mental lightness.
& \textit{``It pulled me out of work without demanding thinking—the lightness made it easy to mentally let go.''} (P7) \\
\cline{2-3}

& Light physical activation supports bodily recovery.
& \textit{``I moved without realizing it; it felt refreshing rather than like deliberate exercise.''} (P1) \\
\cline{2-3}

& Media can provide cognitive rest—or impose cognitive load.
& \textit{``When the content matched my mood, it relaxed me; when it required thinking, it became a bit mentally heavy.''} (P14) \\
\cline{2-3}

& Feedback and interactivity heighten bodily awareness.
& \textit{``Small feedback after each action made me more aware of my body and reminded me that I was actually relaxing.''} (P13) \\
\cline{2-3}

& Balance between stimulation and over-immersion.
& \textit{``If it gets too engaging, I don't want the break to end—it becomes harder to switch back to work.''} (P4) \\
\cline{2-3}

& Completion cues and pacing clarify break boundaries.
& \textit{``The closing sequence and breathing made it clear the break was ending—I felt ready to return to work.''} (P9) \\

\midrule

\multirow{6}{=}{\textbf{Sustained Novelty \& Adaptability}}
& Media variety sustains long-term novelty.
& \textit{``Each session stayed fresh because the content changed—different clips kept me curious every time.''} (P10) \\
\cline{2-3}

& Activity and path variation support repeated use.
& \textit{``Repeating the same path was okay, but having different routes or activities would make long-term use even better.''} (P9) \\
\cline{2-3}

& Fit with personal media preference affects enjoyment.
& \textit{``I don't love talk shows, but with content I enjoy I would definitely use this more often.''} (P15) \\
\cline{2-3}

& Adaptation to physical and social context is crucial.
& \textit{``Some activities depend on where I am—if the system adapts to the real environment, I'd follow more easily.''} (P7) \\
\cline{2-3}

& Temporal adaptability (short vs.\ long breaks).
& \textit{``Short or long breaks both worked—the pacing just needs to match how much energy I have at the moment.''} (P3) \\
\cline{2-3}

& User agency (skip, extend, modify) supports personalization.
& \textit{``Sometimes I wanted to skip a part or extend an activity—having that control would make the experience feel more mine.''} (P12) \\

\bottomrule
\end{tabularx}
\end{table*}

\end{document}